\documentclass{webofc}

\usepackage[T1]{fontenc}
\usepackage{lmodern}
\usepackage[utf8]{inputenc}

\usepackage{dsfont}
\usepackage{amsthm, amssymb, amstext, amsmath, bbm}
\usepackage{float}
\usepackage{tikz}
\usepackage[export]{adjustbox}

\usepackage{subcaption}

\usepackage{algorithm}
\usepackage{algpseudocode}

\usepackage{epstopdf}
\usepackage{blkarray} 

\usepackage[colorlinks=true,
							pdfstartview=FitV, 
							linkcolor=blue,  
							citecolor=blue, 
							urlcolor=blue,
							linktocpage=true]{hyperref}
\usepackage{soul}
\usepackage{braket}

\usepackage[shortlabels]{enumitem}

\usepackage{minitoc}

\renewcommand{\qedsymbol}{$\blacksquare$}

\newcommand{\ii}{{\rm i}}
\newcommand{\Tr}[1]{\mathrm{Tr}\!\left[#1\right]}
\newcommand{\pTr}[2]{\mathrm{Tr}_{#1}\!\left[#2\right]}

\newcommand{\supmat}[1]{\bar{\bar{#1}}}

\newcommand{\sss}{\hat{\rho}_{ss}}
\newcommand{\eig}[1]{\hat{\rho}_{#1}}

\newcommand{\LL}{\mathcal{L}}
\newcommand{\DD}{\mathcal{D}}
\newcommand{\Lmat}{\bar{\bar{\mathcal{L}}}}

\newcommand{\leig}[1]{\hat{\sigma}_{#1}}

\makeatletter
\newcommand*\bigcdot{\mathpalette\bigcdot@{.5}}
\newcommand*\bigcdot@[2]{\mathbin{\vcenter{\hbox{\scalebox{#2}{$\m@th#1\bullet$}}}}}
\makeatother

\renewcommand{\Re}[1]{\mathbb{R}\mathrm{e}\left[#1\right]}

\newcommand{\abs}[1]{\lvert #1 \rvert}

\newcommand{\de}{{\rm d}}

\newcommand{\rhot}{\hat{\rho}(t)}

\newcommand{\expect}[3]{\left\langle #1\middle| #2 \middle| #3 \right\rangle}

\newcommand{\commutator}[2]{\left[#1, #2\right]}
\newcommand{\anticommutator}[2]{\left\{#1, #2\right\}}
\newcommand{\jump}{\hat{\Gamma}}

\renewcommand{\eqref}[1]{Eq.~(\ref{#1})}

\def\be{\begin{equation}}
\def\ee{\end{equation}}
\def\ba{\begin{eqnarray}}
\def\ea{\end{eqnarray}}

\newcommand{\ssx}{\hat{\sigma}^x}
\newcommand{\ssy}{\hat{\sigma}^y}
\newcommand{\ssz}{\hat{\sigma}^z}

\title{Open quantum systems}
\subtitle{A brief introduction}
\author{\firstname{Fabrizio} \lastname{Minganti} \inst{1} \fnsep \thanks{\email{fabrizio.minganti@gmail.com}}
\footnote{Presently at Alice \& Bob.}
\and \firstname{Alberto} \lastname{Biella} \inst{2} \fnsep \thanks{\email{alberto.biella@cnr.it}}}

\institute{{Institute of Physics, \'Ecole Polytechnique F\'ed\'erale de Lausanne (EPFL), \\ CH-1015 Lausanne, Switzerland
}\and {Pitaevskii BEC Center, CNR-INO and Dipartimento di Fisica, Universit\`a di Trento, I-38123 Trento, Italy}}

 \date{\today}

\begin{document}

\abstract{This text is a short introduction to the physics of driven-dissipative many-body systems, focusing on a few selected topics. 
Beyond its more ``historical'' interest in the study of atomic physics and quantum optics, presently the modeling and studying dissipative phenomena in open quantum systems is pivotal to understanding quantum hardware platforms. While the lack of a thermodynamic potential for these out-of-equilibrium open systems makes it theoretically challenging to investigate their physics, at the same time it allows going beyond the thermodynamic paradigms and investigating new and exotic phenomena. 
We will focus on one of the simplest, yet most effective, descriptions of open quantum systems, namely the (Gorini-Kossakowski-Sudarshan-) Lindblad master equation.
This phenomenological approach describes quantum systems that weakly interact with their surrounding environment. 
Although many of the results derived below will apply to any quantum system, we will focus in particular on bosonic/spin systems.
}

\maketitle
\tableofcontents
\section{Introduction}
\label{Chap:Intro}

\begin{figure}
	\centering
	\includegraphics[width=0.6\textwidth, trim=25.2cm 0.2cm 0.cm 0.cm, clip]{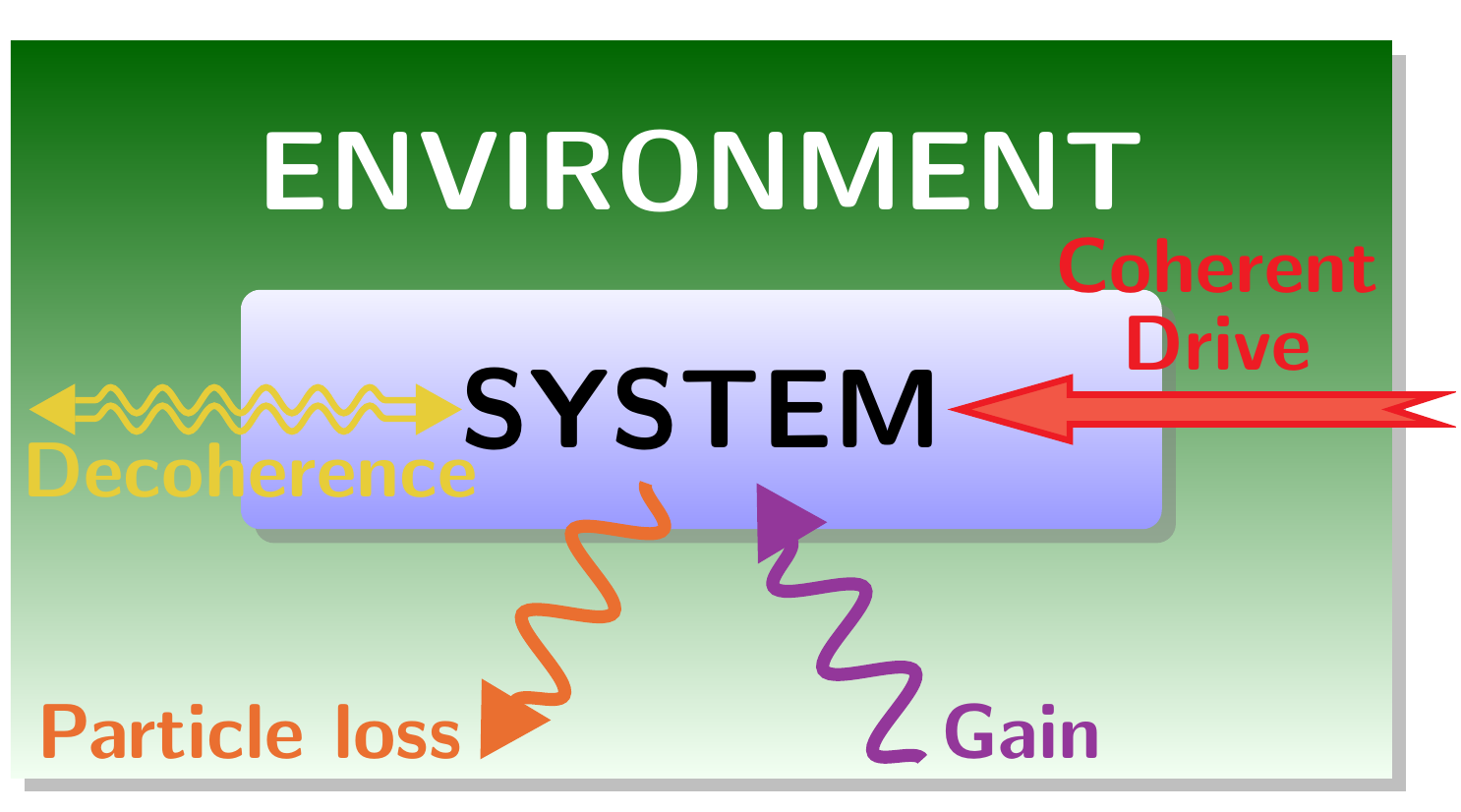}
	\caption{S superconducting circuits, trapped ions, neutral atoms, etc...  can be described in a general fashion as an open quantum system. The system is surrounded by its environment inducing dissipative processes (e.g., decoherence, gain, and particle loss). Moreover, a coherent  (Hamiltonian) drive can continuously repopulate the system.}
	\label{Fig:Scheme}
\end{figure}

The problem of system-environment interaction often emerges in the study of driven  systems, a flourishing research field at the crossroads of condensed matter, quantum optics, and quantum information \cite{CarusottoRMP13,DevoretScience13,NohReview17,kockumNat19}.
In these systems, excitations, energy, and coherence are continuously exchanged with the environment, and they can be driven back via pumping mechanisms \cite{BreuerBookOpen,Wiseman_BOOK_Quantum} (see Fig.~\ref{Fig:Scheme}). 
Hereafter, we will often describe the \textit{bosonic excitations of these systems as photons}.

\subsection{The main ingredients}

As a simple example, let us consider a two-level system (qubit). 
This model can represent the hyperfine microwave transition between two atomic levels in Rydberg atoms, or a photon inside a box at zero temperature \cite{Haroche_BOOK_Quantum}.
If it were to be isolated, such a system would be described as
\begin{equation}\label{Eq:H_sys_2_lvl}
H_{\rm sys} = \frac{\omega}{2} \hat{\sigma}_{z}.
\end{equation}
Thus, in this very simplified picture, the value of $\omega$ determines entirely the characteristics of our two-level system: once we initialize the initial state $\ket{\Psi(t=0)}$, the two-level system will oscillate at a frequency $\omega$ and between states determined by the eigenvectors $\ket{\uparrow}$ and $\ket{\downarrow}$.

No matter the care in isolating a quantum system, it will always interact, even if so slightly, with the environment surrounding it. 
For instance, an excitation may be lost from the two-level system to the environment, in a process known as spontaneous emission.
An initially excited two-level system decays through, e.g., the emission of a photon into the continuum of electromagnetic modes constituting the environment.
Many (more or less quantum) light-emitting devices work on this simple principle of spontaneous emission, from lasers to common light bulbs.

This description of an open quantum system whose dynamics is interrupted by 
discontinuous
events is reminiscent of the ``quantum jumps'' or ``wave function collapses'' of the De Broglie theory of quantum mechanics. 
During these quantum jumps, quantum interferences are lost, and the system seems to evolve under the action of non-unitary operators.
As we will briefly discuss below, these quantum jumps are reminiscent of the problem of wave function collapse upon quantum measurement, and, in standard formulations of quantum mechanics, they can be introduced through the postulates of quantum measurement.

Despite being a self-consistent description of quantum mechanics, this textbook formulation contains a paradoxical argument.
On the one hand, if quantum mechanics is the fundamental description of reality any system must be described using only microscopic quantum mechanical principles. On the other, we assign to the measurement instrument (and thus to the environment) a peculiar character, whose action goes beyond that of quantum mechanics. 
Thus, the paradox: if the measurement instrument is ultimately made of quantum particles, how can quantum mechanical objects induce an effect that is not described by quantum mechanics?

\subsubsection{``Irreversible'' quantum mechanics: a simple numerical benchmark}

The problem of quantum measurement is still debated, with several  theories providing different philosophical solutions to the problem.

A simple argument for the emergence of non-unitary dynamics from unitary interaction can be obtained by considering the following simplified configuration.
Let us go back to our two-level system, and let us suppose that it is coupled to $M$ bosonic modes of energy $\omega_i$ (the environment) of the form:
\begin{equation}\label{Eq:H_env_2_lvl}
H_{\rm env} =  \sum_{i=1}^{M} \omega_i \hat{b}^{\dagger}_i\hat{b}_i .
\end{equation}
Furthermore, we will suppose that the system and environment interact as
\begin{equation}\label{Eq:H_int_2_lvl}
H_{\rm int} =   \sum_{i=1}^{M} g_i(M) \left(\hat{a}^{\dagger}+\hat{a} \right)  \left(\hat{b}^{\dagger}_i+\hat{b}_i \right),
\end{equation}
We will consider that the frequencies $\omega_i$ are normally distributed around the frequency $\omega$ of the qubit. 
This represents the fact that the environment contains many frequencies, whose characteristics change from setup to setup. 
Similarly, we suppose that $g_i(M)$ is normally distributed around a central coupling $\bar{g}(M)$ and that $\bar{g}(M)=\bar{g}(M=1)/M$.
That is, if we increase the number of modes, the coupling decreases (to maintain constant the ``interaction'' between the system and the environment).
All in all, we are treating the environment as a second quantum system, with several random eigenmodes, each one of them weakly and randomly coupled to the original two-level system.

\begin{figure}
    \centering
    \includegraphics[width=0.8 \textwidth]{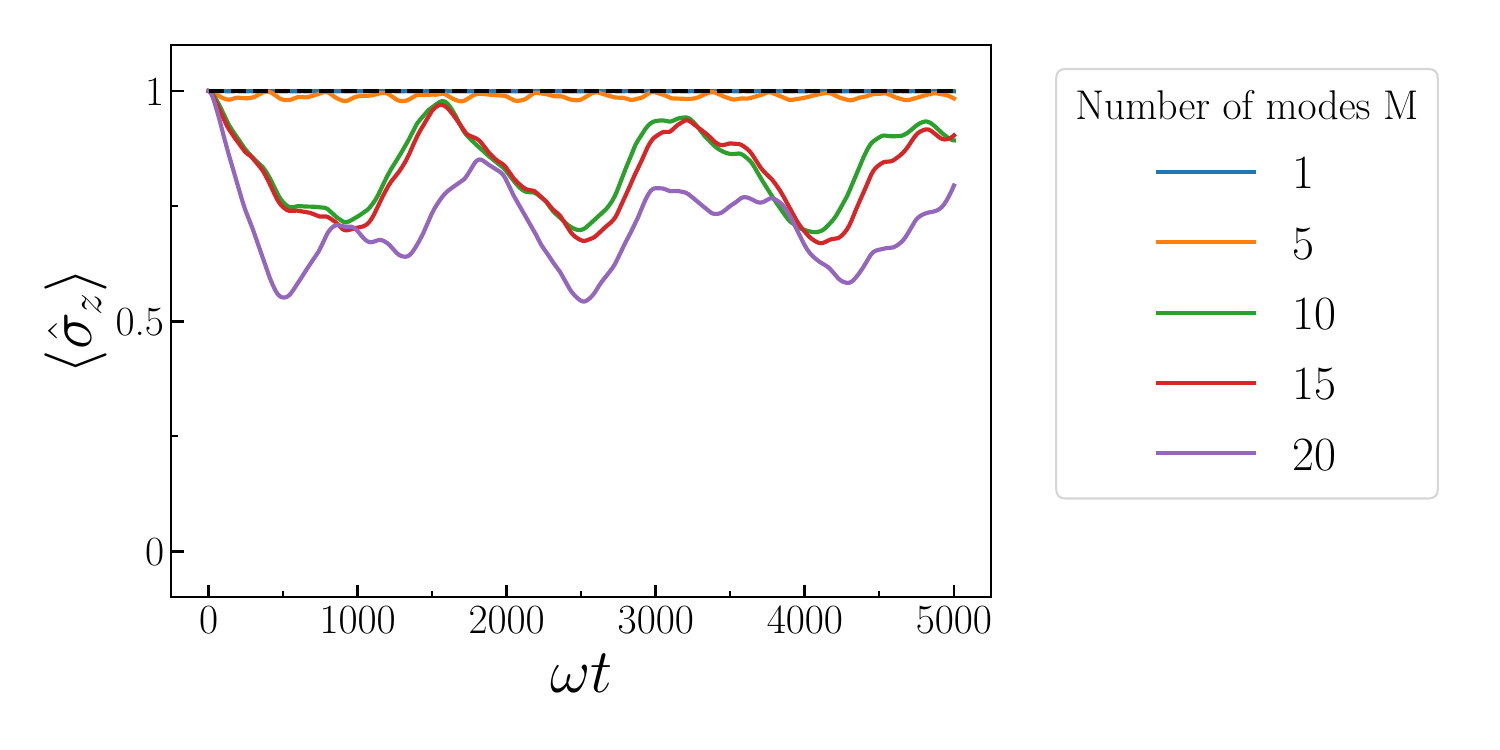}
    \caption{The effect of a random environment on a qubit initialized in $\ket{\uparrow}$ and each of the environment modes in $\ket{0}$ for a different number of modes $M$ and evolving according to \eqref{Eq:Hamiltonian_Qubit_example}. Parameters: $\bar{g}(M=1) = 10^{-3} \cdot \omega$, and the Gaussian distribution of $\omega_i$ and $g_i$ is taken with variance $5\%$.}
    \label{fig:effective_dissipation}
\end{figure}

By combining Eqs.~(\ref{Eq:H_sys_2_lvl}), (\ref{Eq:H_env_2_lvl}), and (\ref{Eq:H_int_2_lvl}) the resulting Hamiltonian is given by
\begin{equation}\label{Eq:Hamiltonian_Qubit_example}
    \hat{H} = \hat{H}_{\rm sys}+ \hat{H}_{\rm env} + \hat{H}_{\rm int}.
\end{equation}
In Fig.~\ref{fig:effective_dissipation} we study the dynamics resulting from $\hat{H}$ as a function of the number of modes.
If the qubit is perfectly isolated (M=0) (black-dashed line) it will not evolve. As soon as we couple to it a single bosonic mode (M=1), we observe that very small Rabi oscillations take place. As we increase the number of qubits, these oscillations become more and more irregular and, eventually, they reach a regime characterized by an initial decay, and the time required for a ``revival'' of the qubit becomes longer and longer. 

All in all, it looks like the qubit is experiencing an initial exponential decay, and then it reaches a ``plateau'' before the excitation lost in the different bosonic modes manages to repopulate the qubit.
In other words, by the effect of destructive interferences, an excitation lost in the environment cannot be re-inject in the qubit. While this simulation has been carried out with only 20 modes, if we were to consider the extremely large number of modes of a real environment, the model would result in the fact that an excitation lost to the environment will never come back to the system.
In other words, by considering (i) many Hamiltonian modes with (ii) a weak coupling, we manage to obtain an ``irreversible'' dynamics from a Hamiltonian one.

It is therefore possible to describe how a quantum system experiences dynamics seemingly described by laws ``outside'' of quantum mechanics.
This behavior resembles that of thermodynamic systems that evolve irreversibly towards their thermal equilibrium due to the coupling to a thermodynamic reservoir.
The stochastic quantum jumps are a phenomenological description of the system-to-environment coupling
derived from fundamental aspects of quantum mechanics.
Such empirically justified dynamical terms can thus be derived from the
well-established quantum mechanics of interacting systems, provided the environment is properly taken into account \cite{BreuerBookOpen}.

Finally, the fascinating point in this interpretation is that it removes the apparent paradox of the measurement in quantum mechanics.
Indeed, from this full quantum mechanical description, it is possible to argue why the reality we observe is classical, a theory known as einselection. Although such a theory goes beyond the purpose of this discussion, we refer the interested reader to Ref.~\cite{ZurekRMP03}.

\subsection{Why open quantum systems?}

\begin{figure}
    \centering
    \includegraphics[width=0.9\textwidth]{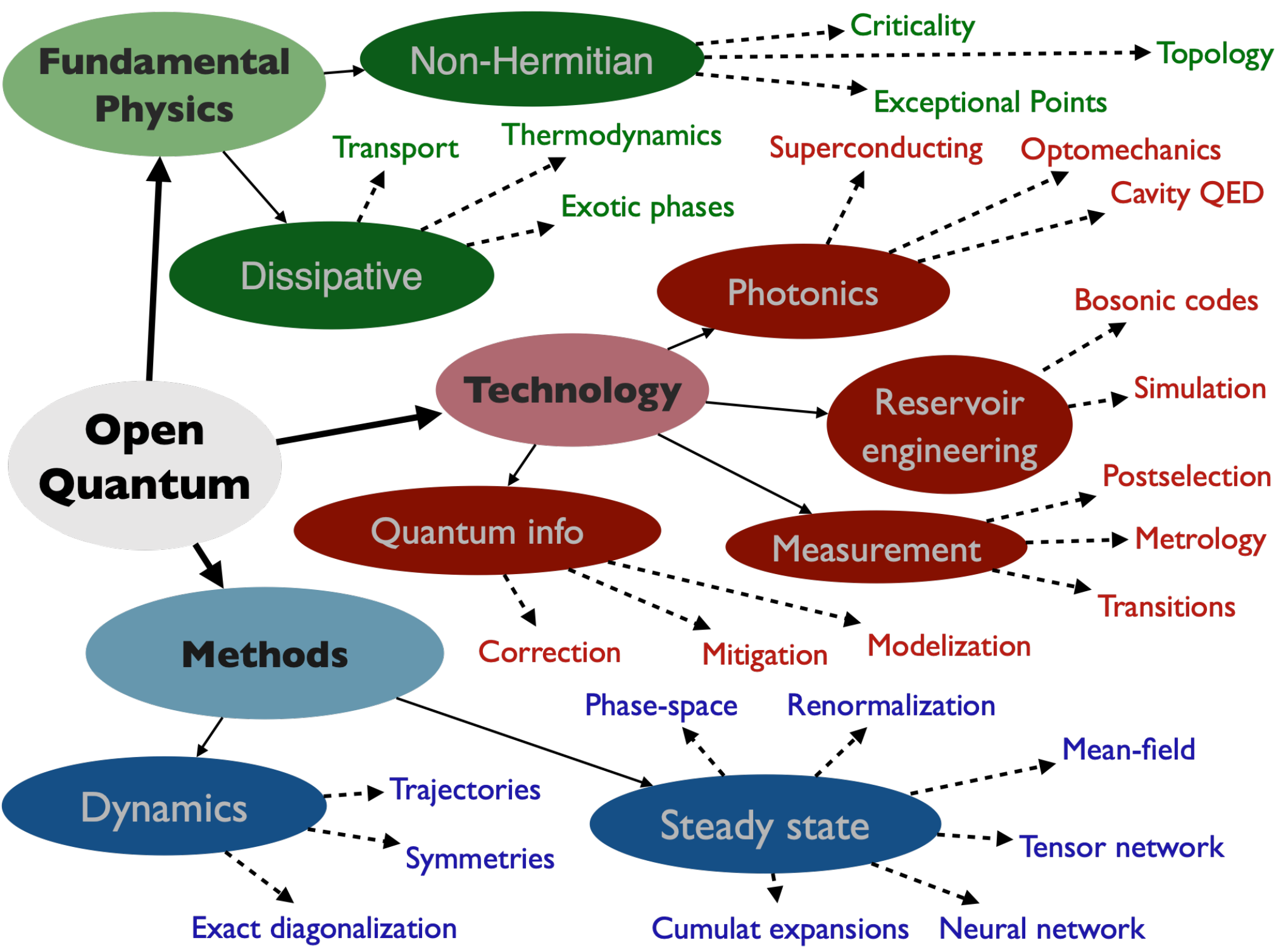}
    \caption{The set of questions in open quantum systems}
    \label{fig:Open_quantum}
\end{figure}

Before entering the details of how to describe open quantum systems, let us briefly motivate \textit{why} it is interesting to study them (see Fig.~\ref{fig:Open_quantum}).

Several experimental platforms -- such as superconducting circuits,
solid-state optical microcavities, optomechanical and atomic systems -- can be modeled within the theoretical formalism of open quantum systems \cite{BrowaeysNatPhys20,LeHurCRP16,AspelmeyerRMP14}. 
These devices, evolving under the action of local and nonlocal Hamiltonian terms, simultaneously being subject to measurement and feedback protocols, stand as a paradigm for quantum technology development.
The same framework can be used to investigate the preservation and manipulation of quantum information.
Furthermore, the field of reservoir engineering \cite{VerstraeteNATPH2009} allows for generating highly-entangled  quantum states.
Finally, there is a close connection between open quantum systems and measurement (as we will also see below). 
For instance, to face the challenge due to the creation and stabilization of macroscopic quantum states techniques that reduce or mitigate the effect of dissipation have to be devised \cite{AruteNat19, SchneiderOptica19,TsePRL19, AndersenNatPhys20,Gil-SantosNat20}. Indeed, if not perfectly protected, the states generated within quantum devices become classical (e.g., a Schr\"odinger cat collapses in its dead or alive state \cite{LeghtasScience15,MingantiSciRep16}).

Dissipation is not only detrimental in quantum technologies but if properly understood and harnessed, dissipative processes can lead to technological advantages. 
The very fact that the physics becomes \textit{non-Hermitian} makes it possible to witness phenomena that, otherwise, would be inaccessible in closed quantum systems.

The physics of an open system is also extremely significant from a fundamental point of view.
Examples include the physics of monitored systems (a special type of open quantum system) and their transitions, as well as the study of the phases of open systems and their transitions.
Examples include boundary (dissipative) time crystals, dissipative phase transition, chaos, lasing, and bistability.

This fundamental and technological interest is accompanied by an effort to determine efficient methods to simulate both the steady state and the dynamics of open quantum systems. 
Indeed, many of the efficient methods that work for closed quantum or thermal systems do not translate into the physics of open quantum systems, making the development of algorithms for an efficient numerical study of many-body-driven-dissipative systems a cutting-edge topic.

\section{Theoretical Framework}
\label{Chap:Framework}

We provide here a very brief description of the theoretical tools needed later to properly describe the open quantum systems.

\subsection{Density matrices, quantum maps, and superoperators}

We consider the rather general problem of a quantum system $S$ coupled to an environment
$E$. 
By environment we mean a large collection of degrees of freedom, each one coupled to the system, with a continuous and wide spectrum of characteristic frequencies, in thermal equilibrium and at some temperature (possibly zero). 
The Hilbert space of $(S \oplus E)$ is $H_{SE}=H_S \otimes H_E$, where $H_S$ is the Hilbert space of the system and $H_E$ is the Hilbert space of the environment.
The environment and system are, together, described by a wave function $\ket{\Psi} \in H_{SE}$.
We are interested in describing $S$, neglecting what happens to $E$.

To do so, one can build the reduced-density operator $\hat{\rho}_S$ by tracing $\hat{\rho}_{SE}=\ket{\Psi}\bra{\Psi}$ over the degrees of freedom of $E$ \cite{Haroche_BOOK_Quantum,Gardiner_BOOK_Quantum,Wiseman_BOOK_Quantum,ParisEPJST2012}, that is:
\begin{equation}
\hat{\rho}_S ={\rm Tr}_E\left[\hat{\rho}_{SE}\right] \equiv
\sum_{k}{\rm Tr} \left[
\left(\ket{\varphi^E_k} \bra{\varphi^E_k}\otimes \hat{\mathds{1}} \right)\,\hat{\rho}_{SE}
\right]
=
\sum_{i,j} c_i c_j \ket{\psi^S_i} \bra{\psi^S_j},
\end{equation}
where
$\ket{\varphi^E_k}$ form a basis of the environment Hilbert space $H_E$, and
$\ket{{\psi}_j^{S}}$ span the system's Hilbert space $H_S$.
The operator $\hat{\rho}\equiv\hat{\rho}_S$  (from now on we will drop the label $S$) contains all the information needed to describe the statistics of outcomes of any measurement performed only on the system.
One is now interested in determining the properties of $\hat{\rho}$.

\subsubsection{Properties of density matrices and ensemble ambiguity theorem}
\label{Sec:Density_matrix}

A density matrix is formally defined as an operator $\hat\rho$ that \cite{Cohen-Tannoudji_BOOK_Quantum_Vol_1,Carmichael_BOOK_1,Gardiner_BOOK_Quantum,Haroche_BOOK_Quantum,ParisEPJST2012,RivasBOOK_Open}
\begin{enumerate}
    \item Has trace 1, i.e., $\Tr{\hat \rho}=1$;
    \item Is Hermitian ($\hat\rho=\hat\rho^\dagger$);
    \item Is postive semi-definite $\expect{\phi}{\hat\rho}{\phi}\geq0$.
\end{enumerate}
These three properties ensure that $\hat \rho$ can be diagonalized, and
\begin{equation}\label{Eq:diagonal_density_matrix}
    \hat \rho = \sum_i p_i \ket{\Psi_i} \bra{\Psi_i}, \quad \sum_i p_i =1, \quad \text{with} \quad p_i>0.
\end{equation}
That is, $p_i$ can be interpreted as the probabilities of mutually excluding events associated with the states $\ket{\psi_i}$.
$\hat \rho$ then represents the general state of a system on which an observer has only partial information, and the probabilities $p_i$ describe the likelihood of finding the system in a certain state upon an appropriate measure.

The diagonalization in \eqref{Eq:diagonal_density_matrix} may be misleading. Indeed, the eigenstates of $\hat\rho$ should not be regarded as ``more important'' but simply as a possible representation of the state of the system. 
For instance, let us consider the state:
\begin{equation}
\hat{\rho}=\frac{\ket{\uparrow_{z}}\bra{\uparrow_{z}} + \ket{\downarrow_{z}}\bra{\downarrow_{z}}}{2},
\end{equation}
where $\ket{\uparrow_{z}}$ ($\ket{\downarrow_{z}}$) is the eigenstate of $\hat{\sigma}_z$ with eigenvalue $+1$ ($-1$).
Such a matrix could be produced by preparing either $\left|\uparrow_{z}\right\rangle$ or $|\downarrow_{z}\rangle$ according to the outcome of a coin toss (which is occurring with probability $1/2$). 
We similarly have
\begin{equation}
\hat{\rho}=\frac{\ket{\uparrow_{x}}\bra{\uparrow_{x}} + \ket{\downarrow_{x}}\bra{\downarrow_{x}}}{2},
\end{equation}
where this time $\ket{\uparrow_{x}}$ ($\ket{\downarrow_{x}}$) is the eigenstate of $\hat{\sigma}_x$ with eigenvalue $+1$ ($-1$).
The preparation procedures are undeniably different. Yet, there is no possible way to tell the difference by making observations of the spin.
In this regard, the true significance of the \textit{eigendecomposition of a density matrix makes sense only upon the definition of how one plans to measure the system}. 

This fact is an instance of the  \textit{ensemble ambiguity theorem}: given any quantum state described by $\hat \rho$, there is no way to tell how it has been prepared and how to interpret it.
The decomposition in eigenstates is useful to capture ``how mixed'' a state is, but the density matrix decomposition is meaningless without discussing how we extract information from $\hat{\rho}$.

\subsubsection{General form of a quantum map}
\label{Sec:Form_of_Maps}
Having introduced the density matrix $\hat{\rho}$, we are now interested in the most general equation for its time evolution.
Such a quantum map $\mathcal{M}$ must transform a density matrix into another, i.e.
\begin{equation}
\hat{\rho}(t+\tau) = \mathcal{M} \rhot.
\end{equation}
Therefore, $\mathcal{M}$ must to satisfy the following properties:
\begin{enumerate}
\item[\emph{P1}:] $\mathcal{M}$ is linear:
\begin{equation}
\mathcal{M}(\alpha \hat{\rho}_1 + \beta \hat{\rho}_2)= \alpha \mathcal{M}( \hat{\rho}_1) + \beta \mathcal{M}( \hat{\rho}_2).
\end{equation}
\item[\emph{P2}:] $\mathcal{M}$ conserves the Hermiticity of $\hat{\rho}$.
\begin{equation}
\left(\mathcal{M} \hat{\rho}\right)^\dagger=\mathcal{M} \hat{\rho}.
\end{equation}
\item[\emph{P3}:] $\mathcal{M}$ conserves the trace
\begin{equation}
\Tr{\mathcal{M} \hat{\rho}}=1.
\end{equation}
\item[\emph{P4}:] $\mathcal{M}$ conserves the complete positivity\footnote{Not only $\mathcal{M}$ maps positive operators to positive operators, but so does the map for $\mathcal{M} \otimes \mathcal{I}$, where $\mathcal{I}$ is the idendity superoperator for an arbitrary second system $S'$.
Indeed, $\mathcal{M} \otimes \mathcal{I}$ is a legitimate quantum map for a system $S \oplus S'$. $\ket{\psi}$ is then an element of this enlarged Hilbert space.}
\begin{equation}
\braket{\psi|\mathcal{M} \hat{\rho}| \psi}\geq 0 \qquad \text{for any }\ket{\psi}.
\end{equation}
\end{enumerate}
The quantum map $\mathcal{M}$ is a \emph{superoperator}, because it acts on operators to produce new operators, just as an operator acts on vectors to produce new vectors \cite{Carmichael_BOOK_2}.

The conditions \emph{P1-P4} limit the structure of the linear superoperator $\mathcal{M}$.
Although we will not show it here, when $\mathcal{M}$ meets these conditions, there exist a set of $\hat{M}_\mu$ operators such that
\begin{equation}
\mathcal{M} \rhot= \sum_{\mu} \hat{M}_\mu \rhot \hat{M}_\mu^\dagger,
\end{equation}
with the normalization condition
\begin{equation}
\label{Eq:Kraus_operators}
\sum_\mu \hat{M}_\mu^\dagger \hat{M}_\mu = \mathds{1}.
\end{equation}
The operators $\hat{M}_{\mu}$ are called Kraus operators, and follow
from Choi's theorem on completely positive maps (for a demonstration, see, e.g., \cite{Haroche_BOOK_Quantum,Barnett_BOOK_Info,ParisEPJST2012}).
The number of Kraus operators is, at most, $N^2$, where $N$ is the dimension of the Hilbert space.
We also stress that Kraus operators need not be unique: any linear unitary transformation mixing them leaves
the quantum map unchanged.
Kraus operators are an extremely powerful tool: indeed, it is possible to compress the infinite complexity of the environment into (an often limited) set of $\hat{M}_\mu$ operators.

\subsubsection{Superoperators and their properties}

In the following, we will often have to deal with superoperators, i.e. linear operators acting on the vector space of operators.
For instance, we can introduce the commutator superoperator, which we define as $\mathcal{A}=\left[\hat{A}, \bigcdot\right] = \hat{A} \bigcdot - \bigcdot \hat{A}$.
With this notation, we mean that $\mathcal{A}$ acting on $\hat{\xi}$ is such that $\mathcal{A} \hat{\xi} = \hat{A} \hat{\xi} -\hat{\xi}\hat{A}$.
The dot simply indicates where the argument of the superoperator is to be placed.
Moreover, we adopt the convention that the action is always on the operator to the immediate right of the dot.
Superoperators can also ``embrace'' their operators, e.g., $\mathcal{A} = \hat{A} \bigcdot \hat{B}$ is such that $\mathcal{A} \hat{\xi} =  \hat{A} \hat{\xi} \hat{B} $.

The dot-notation (i.e., $\bigcdot$) is particularly useful, since it allows for a nesting of superoperators.
For example, consider $\mathcal{A} = \hat{A} \bigcdot \hat{B}$, $\mathcal{C} = \hat{C} \bigcdot \hat{D}$ and $\mathcal{E} = \hat{E} \hat{F} \bigcdot $.
One has,
\begin{equation}
\begin{split}
\mathcal{A} \mathcal{C} & =(\hat{A} \bigcdot \hat{B} )( \hat{C} \bigcdot \hat{D}) =  \hat{A} \hat{C} \bigcdot  \hat{D} \hat{B}, \\
\mathcal{E} \mathcal{A}  & = \hat{E} \hat{F} \hat{A} \bigcdot \hat{B}, \\
\mathcal{A} \mathcal{E} & = \hat{A} \hat{E} \hat{F} \bigcdot \hat{B}.
\end{split}
\end{equation}

Hereafter, we will systematically adopt the following notation: superoperators will be written in calligraphic characters (e.g., $\mathcal{A}$), operators will be denoted by hats (e.g.,  $\hat{A}$),  states and their duals will be expressed in the Dirac notation ($\ket{a}$ and $\bra{a}$). 
Since the operators form a vector space, it is possible to provide a vectorized representation to each element in $H \otimes H$ . For an operator $\hat{A}$, it will be denoted by $\ket{\hat{A}}$.
We choose the convention to convert the matrices into a column vectors as
\begin{equation}
\hat{A}=\begin{pmatrix}
a & b \\
c & d
\end{pmatrix} \longrightarrow \ket{\hat{A}} = \begin{pmatrix}
a \\ b \\ c \\ d
\end{pmatrix}.
\end{equation}
Consequently, to any \emph{linear} superoperator $\mathcal{A}$ it is possible to associate its matrix representation $\bar{\bar{\mathcal{A}}}$.

Since there is no intrinsic definition of inner product in the operator space $H \otimes H$, we introduce the Hilbert-Schmidt one\footnote{That is, given two matrices 
	$$
	\hat{A}=\begin{pmatrix}
	a & b \\
	c & d
	\end{pmatrix}, \quad 	\hat{E}=\begin{pmatrix}
	e & f \\
	g & h
	\end{pmatrix}$$
	one has
	$$\braket{\hat{A}|\hat{E}} = \begin{pmatrix}
	a^* & b^* & c^* & d^*
	\end{pmatrix} \begin{pmatrix}
	e \\ f \\ g \\ h
	\end{pmatrix}=a^* e + b^* f + c^* g + d^*h=\Tr{\hat{A}^\dagger \hat{E}}.$$}:
\begin{equation}
\langle\hat{A}\vert \hat{B}\rangle = \Tr{\hat{A}^\dagger \hat{B}}.
\end{equation}
Hence, the norm of an operator is:
\begin{equation}
\|\hat{A}\|^2=\Tr{\hat{A}^\dagger \hat{A}}.
\end{equation}
Most importantly, having introduced an inner product for the operators, it is possible to introduce the Hermitian adjoint\footnote{There are several different notations in literature to indicate Hermitian conjugation and the symbol $\dagger$ is used with different meanings. In particular, in Ref.~\cite{Carmichael_BOOK_2} the symbol $\hat{A}^\dagger$ indicates a conjugate ``associated'' superoperator.
} of $\mathcal{A}$, which by definition is $\mathcal{A}^\dagger$ such that:
\begin{equation}
\braket{\hat{\xi} | \mathcal{A} \hat{\chi}} = \braket{\mathcal{A}^\dagger \hat{\xi} |  \hat{\chi}}.
\end{equation}
The rules to obtain such adjoint, however, \emph{are not} the same as in the case of operators.
Consider the most general linear superoperator $\mathcal{A} = \hat{A} \bigcdot \hat{B}$.
Exploiting the definition of Hermitian adjoint we have
\begin{equation}
\braket{\hat{\xi} | \mathcal{A} \hat{\chi}} =\Tr{\xi^\dagger \hat{A} \hat{\chi} \hat{B}}= \Tr{\hat{B} \xi^\dagger \hat{A} \hat{\chi}}=
\Tr{(\hat{A}^\dagger \hat{\xi} \hat{B}^\dagger)^\dagger \hat{\chi}} = \Tr{(\mathcal{A}^\dagger \hat{\xi})^\dagger \hat{\chi}}= \braket{\mathcal{A}^\dagger \hat{\xi} | \hat{\chi}}.
\end{equation}
We conclude that
\begin{equation}
\mathcal{A}^\dagger = \hat{A}^\dagger \bigcdot \hat{B}^\dagger.
\end{equation}
We stress that
\begin{equation}
\left(\mathcal{A} \hat{\xi}\right)^\dagger = \left(\hat{A} \hat{\xi} \hat{B}\right)^\dagger = \hat{B}^\dagger \hat{\xi}^\dagger \hat{A}^\dagger \neq \mathcal{A}^\dagger \hat{\xi}^\dagger.
\end{equation}

\subsection{A brief discussion on measurement theory}
\label{Subsec:Measurement}

Since an environment coupled to a quantum system ``extracts'' information from it, the theory of (quantum) measurement is a powerful tool to understand how an open quantum system behaves. 

\subsubsection{Projective measurements}
The more textbook definition of a quantum measurement is the \textit{von Neumann} or \textit{projective} measurement. 
Consider an observable (operator) $\hat{O}$ that is Hermitian.
Since $\hat{O}$ is Hermitian, we can express it using it eigenvalues and eigenvectors as
\begin{equation}\label{Eq:definition_measurement_operator}
    \hat{O} = \sum_j o_j \ket{o_j} \bra{o_j} \qquad \text{ i.e.,} \qquad \hat{O} \ket{o_j} = o_j \ket{o_j}.
\end{equation}
Upon a measurement, a system density matrix $\hat{\rho}$ will collapse onto one of the state 
$\ket{o_j}$ with a probability
\begin{equation}\label{Eq:def_proba_measure}
    p_j = \Tr{\hat{\rho} \ket{o_j} \bra{o_j}}=\Tr{\hat{P}_j \hat{\rho} \hat{P}_j},
\end{equation}
where we have introduced the projectors $\hat{P}_j = \ket{o_j} \bra{o_j}$.
Let us notice that
\begin{enumerate}
    \item $\hat{P}_j$ are Hermitian because $\hat{P}_j^\dagger = \hat{P}_j$;
    \item They are positive because $\hat{P}_j$ has eigenvalue 1;
    \item They are complete because $\sum_j \hat{P}_j = \hat{\mathds{1}}$
    \item They are orthonormal $\Tr{\hat{P}_j \hat{P}_k} = \delta_{j,k}$.
\end{enumerate}

According to these rules, if we measure twice and consecutively the operator $\hat{O}$,
we will obtain both times the same result $o_j$ and $\ket{o_j}$.
Therefore, the density matrix after we read the outcome $o_j$ is $\hat{\rho}'_{j}$ and it reads
\begin{equation}\label{Eq:Read_measure_init}
    \hat{\rho}_{j}' = \frac{\hat{P}_j \hat{\rho} \hat{P}_j}{\Tr{\hat{P}_j \hat{\rho} \hat{P}_j}} \equiv \frac{\mathcal{P}_j \hat{\rho}}{\Tr{\mathcal{P}_j \hat{\rho}}},
\end{equation}
where we introduced the projective measurement superoperator $\mathcal{P}_j = \hat{P}_j \bigcdot \hat{P}_j^\dagger$.
This form explicitly ensures that also $\hat{\rho}_{j}' $ is a well-defined density matrix.

The above result describe the single outcome of a single measure.
However, if we are to repeat the measurement several times on a similarly-initialized system, we would average the outcome by the probability that such a measurement takes place, that is:
\begin{equation}\label{Eq:Unread_Measure_init}
\hat{\rho}'= \sum_j p_j \hat{\rho}_{j}' = \sum_j p_j \frac{\hat{P}_j \hat{\rho} \hat{P}_j}{\Tr{\hat{P}_j \hat{\rho} \hat{P}_j}} = \sum_j \hat{P}_j \hat{\rho} \hat{P}_j = \sum_j  \mathcal{P}_j \hat{\rho},
\end{equation}
where the latter follows from the definition of $p_j$ in \eqref{Eq:def_proba_measure}.

The average result after a measurement in \eqref{Eq:Unread_Measure_init} is identical to the fact that a measurement takes place, but we do not know the outcome of the measure.
The measurement has changed the state of a quantum system, even if we do not know the measurement result. 
The difference between the density operators in Eqs.~(\ref{Eq:Read_measure_init}) and (\ref{Eq:Unread_Measure_init}) can be seen as the lack of the observer's knowledge on the measurement outcome, rather then the system being in a different state.

\subsubsection{POVMs}

That of von Neumann measurement is an idealization of measurement protocols.
Many real measurement instrument, however,  cannot be described through projection operators.
Consider a system containing a photon.
If we want to measure the photon number in the system, according to von Neumann measurement, the measurement protocols would be described by the projection operators
\begin{equation}
\hat{P}_{0}=|0\rangle\langle 0|, \quad
\hat{P}_{1}=|1\rangle\langle 1|.
\end{equation}
This, however, does not correspond to an actual photodetector. A true photodetector ``captures'' a photon and transforms it into a current via a complex series of conversion and amplification processes, and it is the macroscopic current which is eventually measured.
Importantly, if a photon is measured, and we repeat the measurement, we will read that the system contains no photons.
As after the measurement the syste must be in the state $\hat{\rho}' = \ket{0}\bra{0}$,
an educated guess for the form of the measurement operators is 
\begin{equation} \label{Eq:photodetector_Kraus}
\hat{M}_{0}=|0\rangle\langle 0|, \quad
\hat{M}_{1}=|0\rangle\langle 1|.
\end{equation}

We thus need to extend the definition of von Neumann measurement to include for these more general measures.
The framework to discuss generalised quantum measurement is provided by the formalism of positive-operator valued measures (POVM)\cite{Barnett_BOOK_Info,Wiseman_BOOK_Quantum}. 
Consider a measure whose outcomes $r$ are associated to a measure operator $\hat{M}_r$.
Upon measuring the result $r$, the density matrix is modified as
\begin{equation}\label{Eq:Read_Measure}
\hat{\rho}_r'= \frac{\hat{M}_r \hat{\rho} \hat{M}_r^\dagger}{\Tr{\hat{M}_r \hat{\rho} \hat{M}_r^\dagger}} \equiv \frac{\mathcal{M}_r \hat{\rho}}{\Tr{\mathcal{M}_r \hat{\rho}}},
\end{equation}
that is, the density matrix $\hat{\rho}_r$ is obtained by ``projecting'' $\hat{\rho}$ onto the measure operators associated to the outcome $r$.
In this regard, the superoperator $\mathcal{M}_r =\hat{M}_r  \bigcdot \hat{M}_r^\dagger $ describe the measure process having obtained the result $r$.

To describe a probability outcome, we need that $p_r=\Tr{\hat{M}_r \hat{\rho} \hat{M}_r^\dagger}>0$. 
Thus, we conclude that $\hat{M}_r$ is a positive semi-definite operator, and therefore
\begin{equation}
     \expect{\psi}{\mathcal{M}_r \rho}{\psi} \geq 0.
 \end{equation}
In order to assure $\sum_r p_r=1$, we must require
\begin{equation}
\sum_r \hat{M}_r^\dagger \hat{M}_r = \mathds{1}.
\end{equation}
The POVM formalism generalizes projective measures formalism \cite{Barnett_BOOK_Info,Wiseman_BOOK_Quantum}, allowing for a description of measurements associated to non self-adjoint operators.
One can easily verify that the results in \eqref{Eq:photodetector_Kraus} satisfy these conditions.

Let us now consider again the effect of averaging, or of an unread measurement.
Similarly to the previous case, we can define
\begin{equation}
    \hat{\rho}' = \sum_r p_r \mathcal{M}_r \hat \rho = \sum_r \hat{M}_r^\dagger \hat \rho \hat{M}_r \equiv \mathcal{M} \hat \rho.
\end{equation}
What are the properties of  $\mathcal{M}$? (1) $\mathcal{M}$ is linear. (2) Since $(\hat{M}_r^\dagger \hat \rho \hat{M}_r)^\dagger = \hat{M}_r^\dagger \hat \rho \hat{M}_r$ it preserves Hermiticity. (3) Since $\sum_r \hat{M}_r^\dagger \hat{M}_r = \mathds{1}$ it is trace-preserving. (4) Since it is a positive sum of positive semi-definite operator, it is positive-semidefinite.
Remarkably, these are exactly $P1-P4$ of a well-defined quantum map in Sec.~\ref{Sec:Form_of_Maps}. 
In other words, any unread or average measurement defines a quantum map and, vice versa, any quantum map can be defined as the action of a generalized measurement protocol.

\section{The Lindblad master equation and the Liouvillian}
\label{sec:Lindblad}

Using an approach based on measurement theory, and within a few assumptions, we derive here the (Gorini–Kossakowski–Sudarshan–)Linblad master equation \cite{GoriniJMP76,LindbladCMP76} to describe a quantum system coupled to an environment, neglecting the details about the system-environment interaction.
We will then introduce the Liouvillian superopertor, i.e., the generator of the dissipative dynamics, and discuss its properties.
The following discussion was largely inspired by Refs.~\cite{Carmichael_BOOK_1,Carmichael_BOOK_2,Haroche_BOOK_Quantum,Wiseman_BOOK_Quantum,Gardiner_BOOK_Quantum,BreuerBookOpen}.

\subsection{The evolution of the reduced density matrix}

As discussed in Sec.~\ref{Sec:Density_matrix}, a quantum system is described by its density matrix $\hat{\rho}$.
Such a density matrix can describe the system's unitary and non-unitary dynamics, emerging by tracing out and neglecting the information stored in the environment. The question is now what is the form of the quantum map describing the evolution of a quantum system in contact with its environment?

\subsubsection{From quantum map to a differential equation}
\label{subsec:LindbladForm}
Having provided the general properties of any quantum  map $\mathcal{M}$, we are now interested in the specific map describing the time evolution of the density matrix $\hat{\rho}$ of a system weakly coupled to its environment.
First, we make the hypotheses that the environment is much ``bigger'' than the system, and therefore at all times it remains ``close'' to its equilibrium.
That is, the system does not change the properties of the environment (\emph{Born approximation}).
We also assume that the Kraus operators are time independent.
The density operator will evolve (smoothly) as 
\begin{equation}
\label{Eq:Krauss_map_Lindblad}
\hat{\rho}(t+\tau) = \mathcal{M} \rhot = \sum_{\mu} \hat{M}_\mu^\dagger \rhot \hat{M}_\mu = \rhot + \tau \frac{\de \rhot}{\de t} + \mathcal{O}(\tau^2).
\end{equation}
If mathematically  the time interval $\tau$  must be ``infinitesimal'', from a physicist perspective it should be handled with some care.
It must be small at the scale of the system dynamics $\Delta t_S$ (i.e. small compared to all characteristic timescales and relaxation times), so that the modification of $\hat{\rho}$ is only incremental. 
On the other hand, it must be much longer than the correlation time of the environment $\Delta t_E$, so that there are no remaining coherent effects in the system-reservoir interaction.
In the limit in which $\Delta t_E \ll \Delta t_S$, the environment can be thought as memoryless, and we can assume that it is always in its thermal-equilibrium state, i.e., disentangled from the system.
In other words, we are assuming that the environment is a \emph{Markovian} bath for the system.

By considering the appropriate limit $\tau\to 0$, one can arrange the Kraus operators in  \eqref{Eq:Krauss_map_Lindblad} so that one is of the order of unity, while all the others are of order $\sqrt{\tau}$: \footnote{Indeed, one need a part of the quantum map $\mathcal{M}$ which is proportional to the identity, and the rest which is proportional to $\tau$. Therefore, the most general form allowed for the Kraus operator is the one in \eqref{Eq:Generalized_projector_Lindblad}.}
\begin{equation}
\label{Eq:Generalized_projector_Lindblad}
\left\lbrace \begin{split}
\hat{M}_0 &= \mathds{1} - \ii \hat{K} \tau \\
\hat{M}_\mu &=  \sqrt{\tau} \jump_\mu \qquad \text{for } \mu\neq0
\end{split} \right.
\end{equation}
The operator $\hat{K}$ can be split in an Hermitian part, $\hat{I} = (\hat{K}+\hat{K}^\dagger)/2$, and an anti-hermitian one $\hat{J}=\ii (\hat{K}-\hat{K}^\dagger)/2$, so that $\hat{K} = \hat{I} - \ii \hat{J}$.
To the first order in $\tau$ one finds
\begin{equation}
\hat{M}_0^\dagger \rhot \hat{M}_0 = \rhot - \ii \tau \commutator{\hat{I}}{\rhot} - \tau \anticommutator{\hat{J}}{\rhot} + \mathcal{O}(\tau^2),
\end{equation}
where $\commutator{\bigcdot}{\bigcdot}$ indicates the commutator and $\anticommutator{\bigcdot}{\bigcdot}$ is the anticommutator.
By using the normalisation condition of the Kraus operators, one has
\begin{equation}
\mathds{1} = \sum_\mu \hat{M}^\dagger_\mu \hat{M}_\mu = \hat{M}^\dagger_0 \hat{M}_0 + \tau \sum_{\mu \neq 0} \jump^\dagger_\mu \jump_\mu+ \mathcal{O}(\tau^2) = \mathds{1} - 2 \tau \hat{J} +\tau \sum_{\mu \neq 0} \jump^\dagger_\mu \jump_\mu  + \mathcal{O}(\tau^2).
\end{equation}
Considering limit $\tau \to 0$, we conclude $\hat{J} =\sum_{\mu \neq 0} \jump^\dagger_\mu \jump_\mu/2 $.
Therefore, the dynamics of $\rhot$ is dictated by an equation of the form
\begin{equation}
\label{Eq:Lindblad_No_assumptions}
\hat{\rho}(t+\tau) = \rhot - \ii \tau \commutator{\hat{I}}{\rhot} + \tau \sum_{\mu \neq 0} \left(\jump_\mu \rhot \jump^\dagger_\mu - \frac{1}{2} \jump^\dagger_\mu \jump_\mu \rhot - \frac{1}{2} \rhot \jump^\dagger_\mu \jump_\mu\right).
\end{equation}
\eqref{Eq:Lindblad_No_assumptions} means that $\hat{\rho}(t)$ evolves smoothly in time under the action of a completely-positive and trace-preserving linear map. 
Indeed, no assumptions have been made about the nature of the operators $\hat{I}$ and $\jump_\mu$.

To grasp their meaning, however, we can use the analogy with the isolated systems.
In this case, the time evolution is dictated by
\begin{equation}
\label{Eq:Liouvillia-Von-Neumann}
\hat{\rho}(t+\tau) = \rhot -\ii \tau \commutator{\hat{H}}{\rhot},
\end{equation}
where $\hat{H}$ is the Hamiltonian of the isolated system.
Comparing Eqs.~(\ref{Eq:Liouvillia-Von-Neumann}) and (\ref{Eq:Lindblad_No_assumptions}), it is clear that $\hat{I}$ plays the role of a coherent Hamiltonian evolution.
As it will be clarified in the next chapter, $\jump_{\mu}$ are called jump operators and describe the coupling with the environment.
If there are no coherent process linking the environment to the system and their interaction is purely dissipative, one has $\hat{I}=\hat{H}$.
We finally obtain the master equation in the Lindblad form \cite{Carmichael_BOOK_1,BreuerBookOpen,Haroche_BOOK_Quantum,Walls_BOOK_quantum,Gardiner_BOOK_Quantum,Wiseman_BOOK_Quantum}
\begin{equation}
\label{Eq:Lindblad_Master_Equation}
\frac{\partial \rhot}{\partial t} = -\ii \commutator{\hat{H}}{\rhot} + \sum_{\mu \neq 0} \left(\jump_\mu \rhot \jump^\dagger_\mu - \frac{1}{2} \jump^\dagger_\mu \jump_\mu \rhot - \frac{1}{2} \rhot \jump^\dagger_\mu \jump_\mu\right).
\end{equation}

\subsubsection{The driving}
In the previous discussion, we made the hypotheses that the system is weakly coupled to the environment, and that the infinitely-many degrees of freedom of the environment remain unchanged by the system.
If, instead, there exists a part of the environment which coherently exchanges excitations with the system, it will result in an additional effective term in the Hamiltonian.

For example, we can include a coherent driving term representing the excitation of the cavity mode by an external laser of frequency $\omega_0$.
This coupling is described via 
\begin{equation}
\hat{H}_{\rm drive} = g(\omega_0) (\hat{a}^\dagger \hat{d} + \hat{a} \hat{d}^\dagger),
\end{equation}
where $\hat{d}$ is the annihilation operator of the Laser field at a frequency $\omega_0$.
If we suppose that the environment always remains in the coherent state $\ket{\beta}$ of the $\hat{d}$ operator\footnote{The coherent state $\ket{\beta}$ can be defined as that state for which $\hat{d} \ket{\beta}=\beta \ket{\beta}$.}, by partial tracing over the environment we have
\begin{equation}
\hat{H}_{\rm drive} = F \hat{a}^\dagger +F^* \hat{a},
\end{equation}
where $F=g \beta$.
In this regard, we will often encounter Hamiltonian operators describing a coherent drive.

More general drives can be introduced, as long as $\hat{H}_{\rm drive}$ is weak
compared to the system Hamiltonian $\hat{H}$.
Indeed, if the drive is too strong, the Born and Markov approximations, necessary to obtain \eqref{Eq:Lindblad_Master_Equation}, may not be valid.

\subsection{The Liouvillian superoperator}
As we previously said, the Lindblad master equation~\eqref{Eq:Lindblad_Master_Equation} is  linear in $\rhot$. 
Hence, it is possible to associate to it the so-called Liouvillian superoperator $\LL$, defined via
\begin{equation}\label{eq:liouvillian}
\partial_t \rhot=\LL \rhot = - \ii \left[\hat{H}, \rhot\right] + \sum_{\mu\neq 0} \mathcal{D}[\jump_\mu] \rhot,
\end{equation}
where $\mathcal{D}[\jump_\mu]$ is the dissipator, acting as
\begin{equation}
\mathcal{D}[\jump_\mu] \bigcdot= \jump_\mu \bigcdot \jump_\mu^\dagger - \frac{1}{2} \jump_\mu^\dagger \jump_\mu \bigcdot - \frac{1}{2} \bigcdot \jump_\mu^\dagger \jump_\mu.
\end{equation}
The evolution superoperator $\exp(\LL t)$ is trace-preserving and generates a completely positive map since it is associated with the Lindblad master equation.
Accordingly, the formal solution of \eqref{Eq:Lindblad_Master_Equation} is
$\rhot= e^{\LL t} \hat{\rho}(0)$, for an initial condition $\hat{\rho}(0)$ \cite{Carmichael_BOOK_1,Gardiner_BOOK_Quantum,Haroche_BOOK_Quantum,RivasBOOK_Open}.
For a time-independent Liouvillian, there always exists at least one steady state (if the dimension of the Hilbert space is finite \cite{RivasBOOK_Open,BreuerBookOpen}), i.e., a matrix such that 
\begin{equation}
\LL \sss = 0. 
\end{equation}
This equation means that the steady-state density matrix is an eigenmatrix of the superoperator $\LL$ corresponding to the zero eigenvalue.

\subsubsection{The Liouvillian spectrum}

The Liouvillian superoperator is linear, in the sense that 
$\LL (\alpha \hat{\rho}_1 +\beta \hat{\rho}_2  ) = \alpha \LL \hat{\rho}_1 +\beta  \LL  \hat{\rho}_2  $. 
Its spectrum provides information about the evolution of a quantum system.
Indeed, we can define the eigenvalues $\lambda_j$ associated with the eigenmatrices $\eig{j}$ via the relation
\begin{equation}\label{eq:Liouvillian_Eigenstates}
    \LL \eig{j} = \lambda_j \eig{j}.
\end{equation}
Similarly to the Hamiltonian case, where determining the eigenvalues and eigenvectors allows computing the dynamics at any time, we can also determine the state of an arbitrary open system through its eigenoperators and eigenvalues.

In practice, to obtain the eigenspectrum of $\LL$, one often represents $\LL$ as the matrix $\supmat{\LL}$.
The procedure to explicitly obtain $\supmat{\LL}$ is detailed in App.~\ref{sec:LiouvMatr}.
The eigenvalues ${\lambda_i}$ of $\LL$ can be obtained via the resolution of the characteristic equation $\det \left(\Lmat- \lambda_i \mathbb{I} \right) =0$.
Since $\LL$ is not Hermitian, its eigenvectors need not be orthogonal: $\langle\hat{\rho}_i|\hat{\rho}_j\rangle\neq0$.
Notice also that the Liouvillain may not be diagonalizable, as it may display \textit{exceptional points} \cite{MingantiExceptionalPRA}.
For generic Liouvillians, however, these are a manifold of lower codimension in parameter space.

We give, without proof, the following properties of the Liouvillian (see, e.g., Ref.~\cite{MingantPRA18_Spectral}):
\begin{enumerate}
    \item Given \eqref{eq:Liouvillian_Eigenstates}, $e^{\LL t}\eig{i}=e^{\lambda_i t}\eig{i}$ .
    \item $\Tr{\eig{i}}=0$ if $\Re{\lambda_i} \neq 0$.
    \item If $\LL\eig{i}=\lambda_i\eig{i}$ then  $\LL\eig{i}^\dagger=\lambda_i^*\eig{i}^\dagger$. Therefore, if $\eig{i}$ is Hermitian, then $\lambda_i$ has to be real.
    Conversely, if $\lambda_i$ is real and of degeneracy 1,     $\eig{i}$ is Hermitian. More generally, if $\lambda_i$ has geometric multiplicity $n$ and $\LL$ is diagonalizable, it is always possible to construct $n$ Hermitian eigenmatrices of $\LL$ with eigenvalue $\lambda_i$ \footnote{The algebraic multiplicity of $\lambda$ is defined as the number of times $\lambda$ appears as a root of the characteristic equation.
    The geometric multiplicity, instead, is the maximum number of linearly independent eigenvectors associated with $\lambda$.}.
    \item If $\lambda_i=0$ has degeneracy $n$, then there exist $n$ independent eigenvectors of the Liouvillian (the algebraic multiplicity is identical to the geometrical one). Therefore, there exist multiple steady states towards which the system can evolve, depending on the initial condition.
\end{enumerate}

\subsubsection{Spectral decomposition of eigenmatrices}
\label{Sec:Spectral_decomposition}

If the Liouvillian is diagonalizable, we can conveniently use the eigenstates of $\LL$ as a basis of the Liouville space (apart from the exceptional points) \cite{MacieszczakPRL16}.
Under this hypothesis, for any operator $\hat{A}$ there exists a unique decomposition
\begin{equation}
\label{Eq:eigenvectordecomposition}
\hat{A}=\sum_i c_i \eig{i}.
\end{equation}
It can be proved \cite{BreuerBookOpen,RivasBOOK_Open} that $\Re{\lambda_i}\leq 0, \forall i$.
The real part of the eigenvalues is responsible for the relaxation towards the steady-state,
$\sss = \lim\limits_{t \to + \infty} e^{{\mathcal L} t} \rho(0)$.
For convenience, we sort the eigenvalues in such a way that $\abs{\Re{\lambda_0}}<\abs{\Re{\lambda_1}} < \ldots < \abs{\Re{\lambda_n}}$. 
From this definition it follows that $\lambda_0=0$ and $\sss=\eig{0}/\Tr{\eig{0}}$ if the steady state is unique.
We can also identify another relevant quantity: the Liouvillian gap $\lambda=\abs{\Re{\lambda_1}}$, which is also called asymptotic decay rate \cite{KesslerPRA12}, determining the slowest relaxation dynamics in the long-time limit. 

Let us consider a system admitting a unique steady state.
To be physical, $\hat{\rho}(t)$ must be a Hermitian, positive-definite matrix, and with trace equal to one.
Hence, we must have (see property 2 above):
\begin{equation}
\label{eq:time_evol}
\hat{\rho}(t)=\frac{\eig{0}}{\Tr{\eig{0}}}+ \sum_{i\neq 0} c_i (t) \eig{i}=\sss + \sum_{i\neq 0} c_i(0) e^{\lambda_i t} \eig{i}.
\end{equation}

The remaining question is how to determine the coefficients $c_i$.
As the Liouvillian is a non-Hermitian matrix, the standard orthogonality relations of quantum mechanics are substituted by 
\begin{equation}
    \Tr{\leig{j}^\dagger \eig{k}} = \delta_{j,k},
\end{equation}
where we have introduced the left Liouvillian eigenoperators through
\begin{equation}
\LL^\dagger \leig{j} = \lambda_j \leig{j}.
\end{equation}
Such a counterintuitive relation can be understood from standard linear algebra.
A transformation that puts the Liouvillian in its diagonal form reads
\begin{equation}
 \mathcal{R}^{-1}\LL \mathcal{R} = \Lambda, \qquad \text{where} \qquad \Lambda={\rm diag}\left( \lambda_0, \lambda_1, \dots \right).
\end{equation}
$\mathcal{R}$ can be interpreted as the matrix whose $j$th column is the $j$th eigenvector, because $\LL \mathcal{R} = \Lambda \mathcal{R}$.
In the same way, $\mathcal{R}^{-1} \LL = \mathcal{R}^{-1} \Lambda$ which, taking the Hermitian conjugate leads to $\LL (\mathcal{R}^{-1})^\dagger =  \Lambda^* (\mathcal{R}^{-1})^\dagger$.
As such, the left eigenmatrices, determined by the rows of $\mathcal{R}^{-1}$, enstablish the usual orthogonality relation $\braket{\leig{j}, \eig{k}} =\delta_{j, k}$.
Thanks to these relations, we finally have 
\begin{equation}
    \Tr{\leig{j}^\dagger \hat{\rho}(0)} = \sum_k  c_k(0) \Tr{\leig{j}^\dagger \eig{k}} = c_j(0).
\end{equation}

\subsection{Symmetries in open quantum systems}

The introduction of the Liouvillian superoperator allows for an easy discussion of the symmetries of open quantum systems.
Consider a unitary superoperator $\mathcal{U}=\hat{V}\bigcdot \hat{V}^{-1}$ (with $\hat{V}$ unitary) \cite{BaumgartnerJPA08}, such that
\begin{equation}
\mathcal{U}^{-1}\mathcal{L}\,\mathcal{U}=\mathcal{L},
\end{equation}
or, equivalently, $[\mathcal{L},\mathcal{U}]=0$. 
It follows that the matrix representations $\supmat{\mathcal{U}}$ of $\mathcal{U}$ and $\supmat{\LL}$ of $\mathcal{L}$ can be simultaneously diagonalised.
The existence of a symmetry means that the Lindblad master equation cannot mix different ``symmetry sectors'' and $\supmat{\LL}$  can be cast in a block-diagonal form:
\begin{equation}
\label{Eq:block_diagonal}
\supmat{\LL}=
\left[
\begin{array}{c c c c}
\supmat{\LL}_{u_0} & 0 & \dots & 0 \\
0& \supmat{\LL}_{u_1} & \dots & 0 \\
\vdots & \vdots& \ddots & \vdots \\
0 & 0 & \dots & \supmat{\LL}_{u_n}
\end{array}
\right].
\end{equation}

Although a discussion on symmetry in open quantum systems goes beyond the purpose of this work  \cite{BucaNPJ2012,AlbertPRA14}, we can show, with two examples, the peculiar nature of symmetries in open quantum system.
First, the fact that the equation of motion show an invariance do not imply the presence of a conserved quantity.
To see this fact, it is sufficient to consider the Liouvillian for a single damped armonic oscillator:
\begin{equation}
    \LL \rhot = -i \omega [\hat{a}^\dagger \hat{a}, \rhot] + \gamma \hat{a} \rhot \hat{a}^\dagger - \gamma \frac{ \hat{a}^\dagger \hat{a} \rhot + \rhot \hat{a}^\dagger \hat{a}}{2}.
\end{equation}
This equation is invariant under the transformation $\hat{a} \to \hat{a} e^{i \phi}$, i.e., it has a $U(1)$ symmetry.
However,
\begin{equation}
    \braket{\hat{a}^\dagger \hat{a}}(t)  = \braket{\hat{a}^\dagger \hat{a}}(t=0) e^{- \gamma t}. 
\end{equation}
In other words, despite the $U(1)$ invariance of the system (which in standard quantum mechanics would imply particle number conservation) in this case the particle number is not conserved.
In the field of open quantum systems, this is called a \textit{weak symmetry}.

Does this mean that there is no relation between symmetries and conserved quantities?
Consider now the following example:
\begin{equation}
    \LL \rhot = -i \omega [\hat{a}^\dagger \hat{a}, \rhot] + \eta \hat{a}^2 \rhot \left(\hat{a}^\dagger\right)^2 - \eta \frac{ \left(\hat{a}^\dagger\right)^2 \hat{a}^2 \rhot + \rhot \left(\hat{a}^\dagger\right)^2 \hat{a}^2}{2}.
\end{equation}
This equation is still invariant under the transformation $\hat{a} \to \hat{a} e^{i \phi}$, i.e., it has a $U(1)$ symmetry.
However, the dissipative term is, this time $\DD[\hat{a}^2]$, and the jump operators itself is invariant under the transformation $\hat{a} \to -\hat{a}$.
In this case, this symmetry constraints more the system's dynamics.
If we are to consider the change of parity along the dynamics -- parity being described by  $\hat{\Pi} = \exp{(i \pi \hat{a}^\dagger \hat{a})}$ -- we obtain 
\begin{equation}
    \partial_t \braket{\hat{\Pi}}(t)  = -i \omega \braket{[\hat{\Pi}, \hat{a}^\dagger \hat{a}]} + \eta \braket{ \left(\hat{a}^\dagger\right)^2 \hat{\Pi}  \hat{a}^2} - \eta \frac{ \braket{ \left(\hat{a}^\dagger\right)^2\hat{a}^2 \hat{\Pi}} +\braket{\hat{\Pi} \left(\hat{a}^\dagger\right)^2\hat{a}^2 }    }{2} =0. 
\end{equation}
In this case, parity is conserved along the dynamics.
Such a result can be generalized in the following form: the presence of a \textit{strong symmetry} $\hat{V}$, such that $[\hat{H}, \hat{V}]=0$ and $[\hat{J}_l, \hat{V}]=0$ for all $l$ ensures the presence of conserved quantities.

\section{Errors and their correction in a Liouvillian framework}

As an example of the advanced use of Liouvillians, we discuss here their connection with quantum information and quantum error correction.

First, consider a qubit, with Hamiltonian 
\begin{equation}
    \hat{H}_0 = \frac{\omega}{2} \hat{\sigma}_z^{(1)} , 
\end{equation}
representing the bare energy of the two-level system.
We can eliminate this energy by passing in the frame rotating at $\omega$, remaining with the bare Hamiltonian $\hat{H} = 0$, representing the idea that ``nothing'' is happening on the quantum bit.
A state $\ket{\psi}$ will not evolve in time.

\begin{figure}
    \centering
    \includegraphics[width = 0.98 \textwidth ]{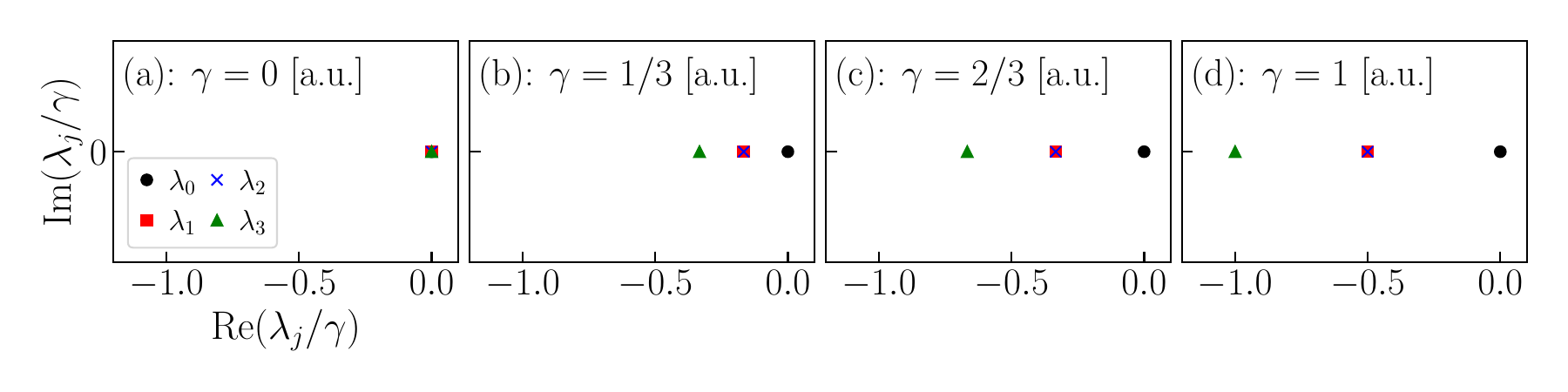}
    \caption{Liouvillian spectrum for a single dissipative qubit, where the dissipation rate is increased between $\gamma = 0$ (a) and $\gamma = 1$ [a.u] (d).}
\label{fig:spectrum_single_qubit}
\end{figure}

The presence of dissipative, however, challenges this idealized picture.
For instance, consider the dissipation of the form
\begin{equation}
    \gamma_1 \mathcal{D}[\hat{\sigma}_{-}].
\end{equation}
What can we now say about the properties of this qubit?

One can study the Liouvillian $\LL = \gamma \mathcal{D}[\hat{\sigma}_-]$ and its spectrum  as a function of $\gamma$, as shown in Fig.~\ref{fig:spectrum_single_qubit}.
For $\gamma = 0$, all the Liouvillian eigenvalues are $\lambda_j =0$.
This fact represents the idea that no matter the initial state, it will not evolve.
For $\gamma \neq 0$, instead, the Liouvillian has the following spectrum:
\begin{subequations} \label{Eq:liouv_spectrum_one_qubit}
\begin{gather}
    \lambda_0 =0, \, \lambda_1 = \lambda_2 = \gamma_1/2, \, \lambda_3 = \gamma_1 .
    \\
    \eig{0} = \ket{0}\bra{0}, \, \eig{1} = \ket{1}\bra{0}, \, \eig{2} = \ket{0}\bra{1}, \, \eig{3} = \ket{1}\bra{1} - \ket{0}\bra{0}.
\end{gather}
\end{subequations}
In this case, the system admits an easy interpretation: namely, any density matrix of the form
\begin{equation}
    \hat{\rho}(0) = \begin{pmatrix}
        a & b \\
        b^* & 1- a
    \end{pmatrix} = \eig{0} + b \eig{1} + b^* \eig{1} + (1 - a) \eig{3}
\end{equation}
will evolve as
\begin{equation}
\label{Eq:}
    \hat{\rho}(t) = \eig{0} + e^{-\gamma t/2} \left(b \eig{1} + b^* \eig{1} \right) + e^{-\gamma t} (1 - a) \eig{3}.
\end{equation}
In other words, these are nothing but the so-called $T_1$ and $T_2$ coherence times of a qubit, and describe the rate at which population and coherences are lost due to the effect of the environment.


\subsection{What makes a logical qubit}

That of a two-level system is often an idealization of more complex multi-level systems.
The question then becomes under which condition one can define within this larger Hilbert space the structure of a logical qubit.

As an example, consider now a larger Hilbert space, for instance, that of two uncoupled qubits.
We are interested in using the first qubit for quantum information purposes, while we will not perform any computational tasks on the second qubit.
For instance, we have
\begin{equation}
    \LL = \gamma_1 \mathcal{D}[\hat{\sigma}_1^{(1)}] +  \gamma_2 \mathcal{D}[\hat{\sigma}_2^{(1)}]
\end{equation}

If we now consider the whole system, we see that its eigenvalues are not $0$, meaning that decay processes will take place.
Obviously, the performance of the second qubit should not be affected by the noise on the first one.
For instance, in the limiting case of $\gamma_1 = 0$, the first qubit is a perfect quantum memory.
So how do we reconcile these two ideas?

If one observes the Liouvillian spectrum in this case, it is of the form
\begin{equation}
\label{Eq:liouv_spectrum_two_qubit} 
\lambda_i^{(1 \& 2)} = \lambda_j^{(1)} + \lambda_k^{(2)}, \quad \eig{i}^{(1 \& 2)} = \eig{j}^{(1)} \otimes \eig{k}^{(2)} 
\end{equation}
where the eigenvalues and eigenoperators are those given in
\eqref{Eq:liouv_spectrum_one_qubit}.
The relevant timescales to describe the loss of logical information are then simply those associated with the eigenvalues 
$\lambda_i^{(1 \& 2)} = \lambda_j^{(1)} + \lambda_0^{(2)}$ and the corresponding states. 
To convince of this, we can again use the eigendecomposition of the Liouvillian to write:
\begin{equation}
\hat{\rho}^{(1) \& (2)}(0) = \sum_{j, k} c_{j, \, k} \eig{j}^{(1)} \otimes \eig{k}^{(2)}
    \quad \to \quad \hat{\rho}^{(1) \& (2)}(t) = \sum_{j, k} c_{j, \, k} e^{(\lambda_j+ \lambda_k) t } \eig{j}^{(1)} \otimes \eig{k}^{(2)}.
\end{equation}
Since we are now interested just in the first qubit, we can trace out the second one.
But as the only object which has nonzero trace is $\hat{\rho}_0$, we finally end up with 
\begin{equation}
\operatorname{Tr}_{2}\left[\hat{\rho}^{(1) \& (2)}(t)\right]  = \sum_{j, k} c_{j, \, k} e^{(\lambda_j+ \lambda_k) t } \eig{j}^{(1)} \otimes \operatorname{Tr}_{2}\left[\eig{k}^{(2)}\right] 
= 
\sum_{j} c_{j, \, k=0} e^{\lambda_j \, t } \eig{j}^{(1)} 
.
\end{equation}
This exactly coincides with the formula for a single qubit.

This simple example demonstrates that the logical qubit  can be ``hidden'' within a larger Hilbert space, and despite the state being possibly very mixed,  for quantum information processing the system can behave exactly like a single qubit.

The fingerprint of the presence of a logical qubit is therefore the fact that the system ``maintains'' four timescales (as a single physical qubit) describing how any point on the logical Bloch sphere evolves.
This idea can be generalized to larger Hilbert spaces, and even mixed states, where the logical qubit is substituted by a logical manifold of states.
In this case, one want to verify that each of these states evolves as if it was one of the state of a single qubit.
This treatment, again, makes use of the theory of Liouvillians and uses techniques similar to those explained in this simple example to define a logical manifold within the Hilbert space.
This is the theory of noiseless subsystems \cite{lidar2013, lidar2014, knill2000, lidar1998, GravinaPRXQuantum,lieu2020} that, however, is far beyond the purpose of this work.

\subsection{The three-qubit repetition code from a Liouvillian perspective}

Error correction codes are essential in quantum computing to mitigate the effects of noise and errors. One such code is the three-qubit repetition code, which provides a simple yet effective method for error correction.

Consider a quantum state encoded using the three-qubit repetition code:
\begin{align*}
\lvert 0_L \rangle &\rightarrow \lvert 000 \rangle \\
\lvert 1_L \rangle &\rightarrow \lvert 111 \rangle
\end{align*}
For the sake of simplicity, assume now that only the operation $\hat{\sigma}_x^{(j)}$ can occur to the system, meaning that 
$\lvert 111 \rangle \to \{ \lvert 110 \rangle, \, \lvert 101 \rangle \, \lvert 110 \rangle \}$ after one dissipative event, and similarly for the $\lvert 000 \rangle$ state.

To detect errors, we measure the state of the qubits using a process called syndrome extraction. The syndrome is a bit string that encodes information about the presence of errors, but that does not degrade the information stored in the qubit.
For example, if we have the state $\lvert \psi_1 \rangle = \alpha \lvert 000 \rangle + \beta \lvert 111 \rangle$, we measure [cf. \eqref{Eq:definition_measurement_operator}]
\begin{equation}
\hat{M}_1 = \hat{\sigma}_z^{(1)} \otimes \hat{\sigma}_z^{(2)}, \quad \hat{M}_2 =\hat{\sigma}_z^{(2)} \otimes \hat{\sigma}_z^{(3)}.
\end{equation}
The outcomes $m_1$ and $m_2$ of $\hat{M}_1$ and $\hat{M}_2$ can take values $\pm 1$.
For the initial state $\lvert \psi_1 \rangle$, $m_1 = m_2 = 1$.

Each time an error occurs, the values of $m_1$ and $m_2 $ change.
We can therefore assume that, after the detection of a single error, we can recover the original state by an appropriate recovery operation (i.e., the action of a Hermitian gate).
For instance, assuming a single-bit flip, we have
\begin{center}
\begin{tabular}{cccccc}
 Bit-flip  & $\left|\psi_1\right\rangle$ & $m_1$ & $m_2$ & Recovery & $\left|\psi_2\right\rangle$ \\
\hline - & $\alpha|000\rangle+\beta|111 \rangle$ & 1 & 1 & $\hat{I} \otimes \hat{I} \otimes \hat{I}$ & $\alpha|000\rangle+\beta|111\rangle$ \\
$\hat{\sigma}_x^{(1)}$ & $\alpha|100\rangle+\beta|011\rangle$ & -1 & 1 & $\hat{\sigma}_x^{(1)}$ & $\alpha|000\rangle+\beta|111\rangle$ \\
$\hat{\sigma}_x^{(2)}$ & $\alpha|010\rangle+\beta|101\rangle$ & -1 & -1 & $\hat{\sigma}_x^{(2)}$ & $\alpha|000\rangle+\beta|111\rangle$ \\
$\hat{\sigma_x^{(3)}}$ & $\alpha|001\rangle+\beta|110\rangle $& 1 & -1 & $\hat{\sigma}_x^{(3)}$ & $\alpha|000\rangle+\beta|111\rangle$
\end{tabular}
\end{center}

\begin{figure}
    \centering  
    \includegraphics[width = 0.78 \textwidth]{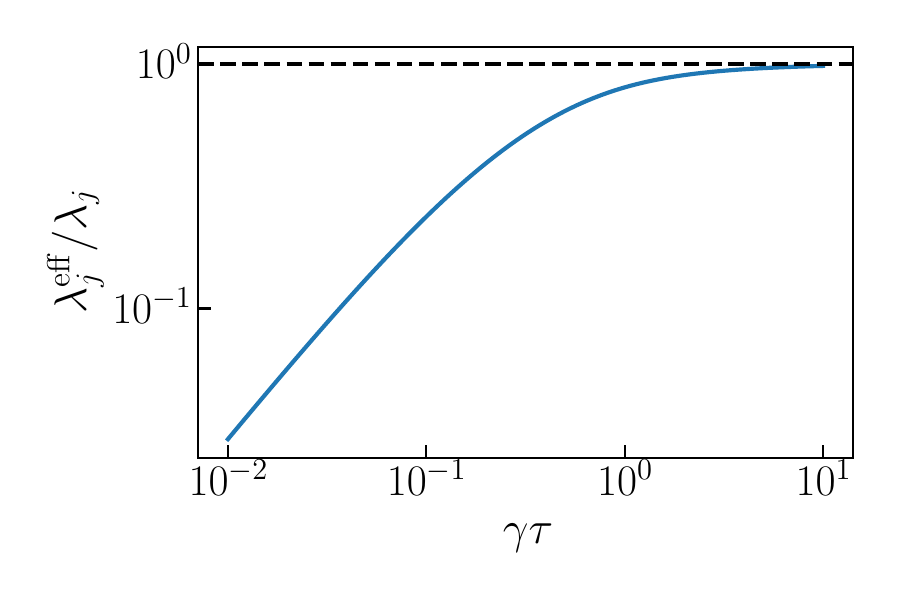}
    \caption{Logicalerror rate in the error corrected code compared to the original one, as a function of the error correction period $\tau$.}
    \label{fig:ratio_error_correction}
\end{figure}

This simplified description gives an idea of the functioning of error correction.
If, however, we want to gauge the performance of an error correction code, we need to consider two additional factors: (1) the rate at which error correction is performed; (2) the fact that errors are not described as the simple action of $\hat{\sigma}_x^{(j)}$, but they require a whole dissipator to describe them.
To this extent, we assume dissipation of the form
\begin{equation}
    \LL = \gamma \left( \DD[\hat{\sigma}_x^{(1)}] + \DD[\hat{\sigma}_x^{(2)}] + \DD[\hat{\sigma}_x^{(3)}]\right).
\end{equation}
We periodically perform our error detection every time $\tau$, and we also assume that the error recovery procedure is instantaneous.
This process is described through the evolution superoperator
\begin{equation}
    \mathcal{E}(\hat{\rho}) = 
    \mathcal{R} e^{\mathcal{L} \tau} \hat{\rho},
\end{equation}
where $\mathcal{R}$ is the detection-and-recovery superoperator.
As the measurement of $\hat{M_1}$ and $\hat{M_2}$ are projective measurement, we can decompose 
\begin{equation}
    \mathcal{R} = \sum_{m_1, \, m_2} \hat{O}_{m_1, \, m_2} \hat{P}_{m_1, \, m_2} \bigcdot \hat{P}_{m_1, \, m_2}^\dagger \hat{O}_{m_1, \, m_2}^\dagger 
\end{equation}
where
\begin{subequations}
\begin{align}
    &\hat{P}_{1, \, 1} = \ket{1,1,1}\bra{1,1,1} + \ket{0,0,0}\bra{0,0,0}  &
     &\hat{O}_{1, \, 1} = \hat{I} \otimes \hat{I} \otimes \hat{I}
    \\ 
    & \hat{P}_{-1, \, 1} = \ket{0,1,1}\bra{0,1,1} + \ket{1,0,0}\bra{1,0,0} &
     &\hat{O}_{-1, \, 1} = \hat{\sigma}_x^{(1)}    \\ 
    & \hat{P}_{-1, \, -1} = \ket{1,0,1}\bra{1,0,1} + \ket{0,1,0}\bra{0,1,0} &
     &\hat{O}_{-1, \, -1} = \hat{\sigma}_x^{(2)}
    \\ 
    & \hat{P}_{1, \, -1} = \ket{1,1,0}\bra{1,1,0} + \ket{0,0,1}\bra{0,0,1} &
     &\hat{O}_{1, \, -1} = \hat{\sigma}_x^{(3)}
\end{align}
\end{subequations}
First, notice that $\sum_{m_1, \, m_2}  \hat{P}_{m_1, \, m_2} = \hat{I} \otimes \hat{I} \otimes \hat{I}$, confirming that they indeed define a projective measurement.
One can then compute, as a function of $\tau$, what are the eigenvalues $\epsilon_j$ of $\mathcal{E}$, to see the performance of the error correction protocol.
In particular, one can define $\lambda_j^{\rm eff} = \log(\epsilon_j)/ \tau$, representing the loss of information per unit of time, and compare it to the eigenvalues of the original non-error corrected Liouvillian.
We plot, for instance, the logical bit-flip error rate in the error-corrected code in Fig.~\ref{fig:ratio_error_correction}.
As one can see, decreasing $\tau$ and thus increasing the error correction rate, improves the performance of the code.

\section{Quantum trajectories}
\label{Chap:Quantum_trajectories}

Having introduced the description of an open quantum system in terms of the Lindblad master equation, we propose here a different interpretation of the physics of open quantum systems.
Indeed, it is not evident the connection between the continuous nonunitary evolution of a Lindblad master equation and the fact that we frequently observe that particles are emitted by a quantum system in random and abrupt events.
The connection between two distinct interpretations can be derived in terms of quantum jumps and quantum trajectories.

\subsection{Dissipation measures the system through quantum jumps}
\label{Sec:Quantum_jumps}

Equipped with the Lindblad master equation and the associated Liouvillian superoperator, as well as their physical interpretation through the coherent Hamiltonian evolution and the effect of dissipation, one still has to determine the form of these quantum jumps $\jump_l$.
To do that, we exploit the close analogy between the Kraus operators and the formalism of generalized measures.

\subsubsection{Connection between quantum jumps and Krauss operators}
As we discussed in the previous chapter (c.f. Sec.~\ref{Subsec:Measurement}), any quantum map can be thought as a set of positive-operator valued measures (POVM)\cite{Barnett_BOOK_Info,Wiseman_BOOK_Quantum}. 
Let us now consider a system in which $\hat{H}=0$ and there exists a unique dissipator $\jump$.
In this case, \eqref{Eq:Generalized_projector_Lindblad} is:
\begin{equation}
\left\lbrace \begin{split}
\hat{M}_0 & = \mathds{1} -  \frac{\tau}{2} \jump^\dagger \jump, \\
\hat{M}_1 & = \sqrt{\tau} \jump,
\end{split} \right.
\end{equation}
and \eqref{Eq:Lindblad_No_assumptions} becomes:
\begin{equation}
\hat{\rho}(t+\tau) = \hat{M}_0 \rhot \hat{M}_0^\dagger + \hat{M}_1 \rhot \hat{M}_1^\dagger = \mathcal{M}_0 \rhot  + \mathcal{M}_1 \rhot ,
\end{equation}
where we recall $\mathcal{M}_\mu = \hat{M}_\mu \bigcdot  \hat{M}_\mu^\dagger$. 
We notice that $\mathcal{M}_0+\mathcal{M}_1$ can be interpreted as measure superoperators since $\hat{M}_\mu$ are POVM.
Moreover, with a probability of order one, the quantum state of the system is unchanged ($\hat{M}_0 \simeq \mathds{1}$) or, with probabilities of order $\tau$, the system undergoes a large evolution (described by $\hat{M}_1 \rhot \hat{M}_1^\dagger$).\footnote{In this sense, the factor $ \tau \jump^\dagger \jump/2$ in $\hat{M}_0$ can be interpreted as the backaction of the measure, introducing a normalisation term needed to preserve the trace of the density matrix.}
We can therefore interpret the Linblad master equation as the the time evolution of a system subject to a \emph{continuous and unread measure} \cite{Haroche_BOOK_Quantum}.

We have, thus, a very simple path to guide us in writing the proper Lindblad master equation for any system,
since this picture can be easily generalized both to a nonzero Hamiltonian and to several $\jump_\mu$.
Accordingly, the  various $\jump_\mu$  are \emph{jump operators} describing a random (perhaps large) evolution of the system which suddenly changes under the influence of the environment. 
Of course, the density matrix $\rhot$ evolves continuously, since the probability of a quantum jumps is finite and proportional to the time step $\tau$.

The interest of the method is that, in many cases, the nature of the quantum jumps can be guessed from the mere nature of the system.
Once again,  we stress that the jump operators are not uniquely defined, since the same relaxation processes can be modelled in different ways, resulting in different unrevealing of the master equation. 
In some situations, the nature of the coupling to the environment may privilege one of these unravellings. For instance, for an atom completely surrounded by a photodetector array, $\jump$ correspond to a photodetection.
As we will see, different unrevealing schemes may correspond to different ways of monitoring the environment (photon counters, homodyne recievers \ldots).
However, all the Lindblad master equations stemming from those different jump operators are equivalent one to the other.

\subsection{The stochastic Schr\"odinger equation}
\label{Sec:SSE}

In Sec.~\ref{Sec:Quantum_jumps} we saw that the structure of the Lindblad master equation admits an interpretation in terms of a continuous and unread measure.
The question now is which kind of evolution does the system undergo if we, instead, keep track of the results of such a measure.

\begin{figure}
\centering
\includegraphics[width=0.9\linewidth]{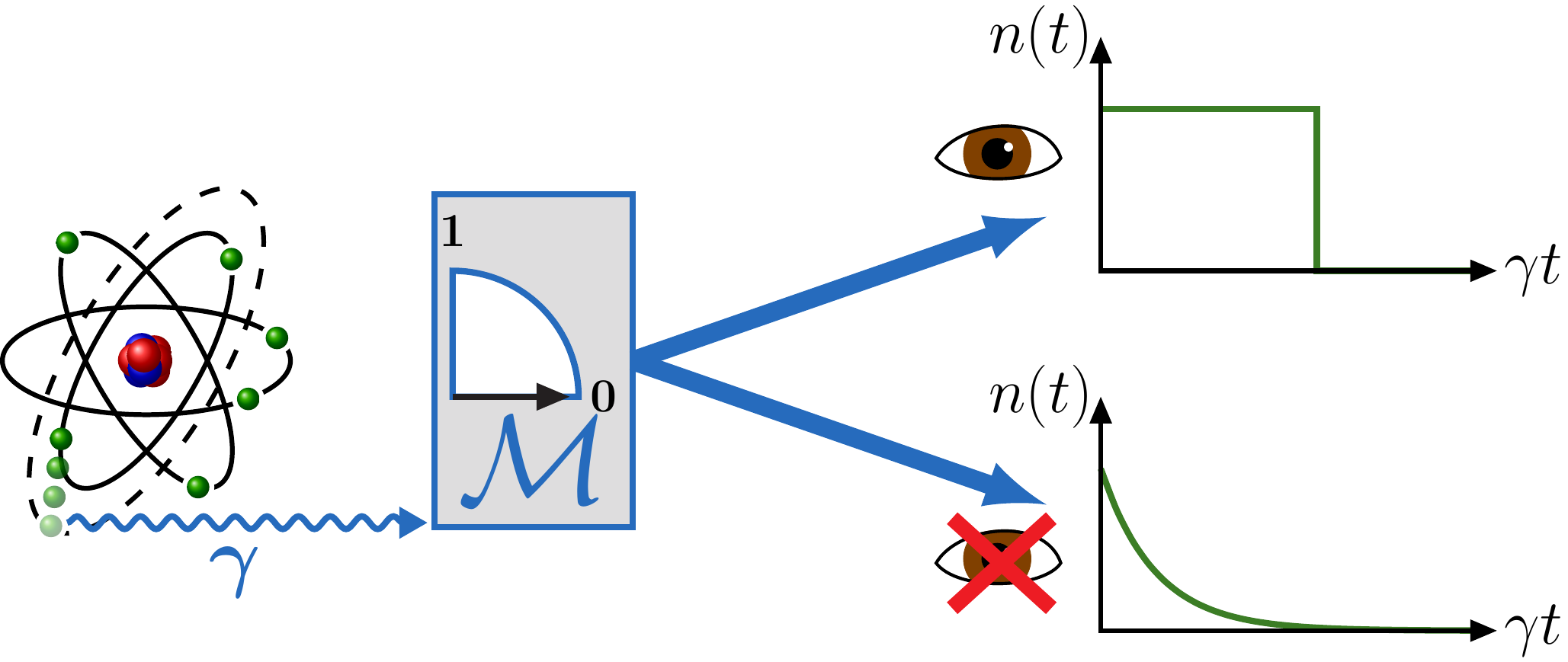}
\caption{Scheme of an apparatus measuring the spontaneous emission of an atom.
The time evolution of $n(t)$ (defined in the main text) is due to a Linblad master equation (unread measure bottom) or to a stochastic process (read measure top).
}
\label{fig:Lindblad_vs_stochastic}
\end{figure}

To explain the qualitative difference between a read and an unread measure procedure, we consider the following example.
An atom is prepared in an excited state $\ket{e}$.
At any time, it may emit a photon into the environment, passing into its ground state $\ket{g}$.
A perfect photodetector immediately signals such an event, thus destroying the photon (c.f. Fig.~\ref{fig:Lindblad_vs_stochastic}).
The quantity $n(t)= \braket{\hat{n}(t)}$, where $\hat{n}=\ket{e}\bra{e}$, indicates whether the system is in its excited or ground state.
If we do not read the photodetector, we provide a probabilistic description of the system, so that $n(t)$ evolves smoothly towards $0$ in accordance to \eqref{Eq:Measure_derivation}.
On the contrary, if the result of the measure is read, we are certain of the state of the atom: either in the excited state or in the ground one.
The time evolution of $n(t)$ will therefore be of stochastic nature: at any moment the photon may be detected, corresponding to an abrupt jump of $n(t)$.
The equation of motion for a system whose environment is continuously and perfectly probed is called a quantum trajectory \cite{Molmer1993,Haroche_BOOK_Quantum,Carmichael_BOOK_2,Wiseman_BOOK_Quantum,DaleyAdvancesinPhysics2014}.

\subsubsection{Jump-counting trajectories}
\label{Sec:Conunting_Trajectories}

Suppose there exist a system whose coherent evolution is dictated by $\hat{H}$, and the environment can be modelled with a single jump operator $\jump$. 
We continuously measure it at a rate $\tau$ with an instrument characterised by two possible outcomes: $r=0$, associated to the superoperator $\mathcal{M}_0$, and $r=1$, associated to $\mathcal{M}_1$. 
We recall $\mathcal{M}_\mu = \hat{M}_\mu \bigcdot  \hat{M}_\mu^\dagger$ and
\begin{equation}
\label{Eq:Measure_Operators_Lindblad_One_Species}
\left\lbrace \begin{split}
\hat{M}_0 & = \mathds{1} -\tau \left( \ii  \hat{H} +  \frac{1}{2} \jump^\dagger \jump\right), \\
\hat{M}_1 & = \sqrt{\tau} \jump.
\end{split} \right.
\end{equation}
The probability to obtain the result $r=1$ is
\begin{equation}
p(t, r=1) = \Tr{\mathcal{M}_1\rhot } = \tau \Tr{\rhot \jump^\dagger \jump},
\end{equation}
and the one to obtain $r=0$ is $p(t, r=0) = 1-p(t, r=1)$.
In the limit $\tau \to 0$, at almost all times the result of the continuous measurement will be $r = 0$, and the system will undergo a smooth (but not unitary) evolution dictated by $\mathcal{M}_0$.
At random times, whose mean rate is $p(t, r=1)$ the system will, instead, experience  a finite evolution due to the jump operator $\jump$ in $\hat{M}_1$.
This abrupt change in $\rhot$ is called a \emph{quantum jump}.

For the sake of simplicity, let us suppose the system is in a pure state $\ket{\psi(t)}$, so that $\rhot= \ket{\psi(t)} \bra{\psi(t)}$ (the generalization to mixed states can be done exploiting the linearity of any quantum map).
To keep track of the continuous measure, let us define the parameter $N(t)$, which counts the number of quantum jumps that have occurred up to the time $t$.
Accordingly, $N(t=0)=0$. 
If a quantum jump occurs between time $t$ and $t+\tau$, the result of the measure is $r=1$ and $N(t+\tau) =N(t)+1$.
Otherwise, if $r=0$, $N(t+\tau) =N(t)$.
Thus, we can introduce the (It\^o) stochastic increment $\de N(t) = N(t + \tau) - N(t)$.
It obeys:
\begin{subequations}
\begin{align}
\de N(t) ^2 & = \de N(t) \\
E[\de N(t)]&= p(t, r=1) = \tau \Tr{\rhot \jump^\dagger \jump} = \tau \braket{\psi(t) | \jump^\dagger \jump |\psi(t)},
\end{align}
\end{subequations}
where $E[\de N(t)]$ indicates the expected value of the stochastic variable $\de N(t)$, and the last equality follows from the hypothesis of a pure state.
From the two previous conditions, it follows also $\tau \de N(t)   = 0 $\footnote{To understand this definition of a stochastic variable, considering $E[\tau \de N(t)]   \propto \tau^2 $ which is infinitesimal in any first-order approximation in $\tau$.}.

Let us recall that a measure whose result is $r$ transforms $\rhot$ as
\begin{equation}\label{Eq:Evolution_Measure_Counting}
\hat{\rho}_r(t+ \tau)= \frac{\hat{M}_r \rhot  \hat{M}_r^\dagger}{\Tr{ \hat{M}_r \rhot \hat{M}_r^\dagger} }=\frac{\hat{M}_r \ket{\psi(t)} \bra{\psi(t)}  \hat{M}_r^\dagger}{{\braket{\psi(t) | \hat{M}_r^\dagger \hat{M}_r |\psi(t)}}},
\end{equation}
where the last follows from the hypothesis of a pure state and the denominator ensures the condition $\Tr{\hat{\rho}(t+ \tau)}=1$.
From \eqref{Eq:Evolution_Measure_Counting} follows that, given an initial pure state, $\rhot$ will always remain pure. 
Therefore, it is sufficient to consider the wave function $\ket{\psi(t)}$ to fully describe the evolution of the system.

The time evolution of $\ket{\psi(t)}$ is, thus, the following. If $\de N(t) = 1$, i.e., a quantum jump happens,
\begin{equation}
\ket{\psi(t+ \tau, \de N(t)=1)}=\frac{\hat{M}_1 \ket{\psi(t)} }{\sqrt{{\braket{\psi(t) | \hat{M}_1^\dagger \hat{M}_1 |\psi(t)}}} } = \frac{\jump \ket{\psi(t)} }{\sqrt{\braket{\jump^\dagger \jump }}},
\end{equation}
where we denoted $\braket{\jump^\dagger \jump }= \braket{\psi(t)|\jump^\dagger \jump | \psi(t)}$.
Otherwise, if there is no detection, $\de N(t)=0$, and 
\begin{equation}
\begin{split}
\ket{\psi(t+ \tau, \de N(t)=0)}& =\frac{\hat{M}_0 \ket{\psi(t)} }{\sqrt{{\braket{\psi(t) | \hat{M}_0^\dagger \hat{M}_0 |\psi(t)}}} }  = 
\frac{\left[\mathds{1} -\tau \left( \ii  \hat{H} -  \frac{1}{2} \jump^\dagger \jump\right)\right] \ket{\psi(t)} }{\sqrt{\braket{\psi(t) |  \mathds{1} -  \tau \jump^\dagger \jump  + \mathcal{O}\left(\tau ^2\right) |\psi(t)}}} \\
& =\left[\mathds{1} -\tau \left( \ii  \hat{H} -  \frac{1}{2} \jump^\dagger \jump\right)\right] \left[\mathds{1} + \tau \frac {\braket{\jump^\dagger \jump }}{2} + \mathcal{O} \left(\tau^2\right)\right] \ket{\psi(t)} \\
&= \left[\mathds{1} - \tau \left(\ii \hat{H} + \frac{\jump^\dagger \jump}{2} -  \frac{\braket{\jump^\dagger \jump }}{2} \right)\right]  \ket{\psi(t)}.
\end{split}
\end{equation}
Finally, one obtains a nonlinear stochastic Schr\"odinger equation of the form
\begin{equation}
\begin{split}
\ket{\psi(t+ \tau)} &= \de N(t) \ket{\psi(t+ \tau, \de N(t)=1)} + \left[1- \de N(t) \right] \ket{\psi(t+ \tau, \de N(t)=0)} \\
 &= \de N(t) \frac{\jump \ket{\psi(t)} }{\sqrt{\braket{\jump^\dagger \jump }}} +  \left[1- \de N(t) \right] \left[\mathds{1} - \tau \left(\ii \hat{H} + \frac{\jump^\dagger \jump}{2} -  \frac{\braket{\jump^\dagger \jump }}{2} \right)\right]  \ket{\psi(t)},
\end{split}
\end{equation}
which in its differential form becomes
\begin{equation}
\label{Eq:SSE}
\de \ket{\psi(t)}= \left[\de N(t) \left(\frac{\jump}{\sqrt{\braket{\jump^\dagger \jump }}} - \mathds{1}\right) + - \ii  \tau \hat{H}_{\rm eff} \right]\ket{\psi(t)},
\end{equation}
where we used $\de N(t) \tau=0$ and we introduced the effective Hamiltonian
\begin{equation}\label{Eq:Effective_Hamiltonian_counting}
\hat{H}_{\rm eff} =\hat{H} - \ii  \frac{\jump^\dagger \jump}{2} +\ii   \frac{\braket{\jump^\dagger \jump }}{2} .
\end{equation}
We will call \eqref{Eq:SSE} a counting stochastic Schr\"odinger equation, since it is a purity-preserving equation which depends on the ``counting'' stochastic parameter $ N(t)$.
We will call a solution to this equation a \emph{counting quantum trajectory} for the system.

\subsubsection{Homodyne trajectories}
\label{Sec:Homodyne_Trajectories}

Clearly, \eqref{Eq:SSE} is not the only possible stochastic Schr\"odinger equation which one can obtain from a Lindblad equation: different choices of the Kraus operators would lead to different evolutions.
Consider a system described by an Hamiltonian $\hat{H}$ and subject to only one jump operator $\jump$.
The following transformation does not modify the structure of the Lindblad master equation
\begin{equation}
\label{Eq:Sustitution_Homodyne}
\jump \to \jump + \beta, \qquad \qquad \hat{H} \to \hat{H} - \frac{\ii\beta}{2} \left( \jump - \jump^\dagger \right) ,
\end{equation}
where $\beta$ is a real number.
Under this transformation the measure operators of \eqref{Eq:Measure_Operators_Lindblad_One_Species} become
\begin{equation}
\label{Eq:Measure_Operators_Lindblad_Homodyne}
\left\lbrace \begin{split}
\hat{M}_0 & = \mathds{1} -\tau \left[ \ii  \hat{H}  +  \frac{\beta}{2} (\jump - \jump^\dagger) +  \frac{1}{2} ( \jump^\dagger + \beta)( \jump + \beta) \right], \\
\hat{M}_1 & = \sqrt{\tau} (\jump + \beta).
\end{split} \right.
\end{equation}
From a physical point of view, the parameter $\beta$ can be thought as a constant coherent field which is continuously measured together with the dissipated particles of our system.

One injects \eqref{Eq:Sustitution_Homodyne} into \eqref{Eq:SSE}, obtaining a stochastic equation of the form
\begin{equation}
\label{Eq:SSE_Homodyne}
\begin{split}
\de \ket{\psi(t)}= &\left[ \de N(t)   \left( \frac{\jump+\beta}{\sqrt{\braket{(\jump^\dagger +\beta) (\jump+\beta)}}}  - \mathds{1}  \right) \right .
\\ & \qquad \left.+ \tau \left(- \ii \hat{H} - \beta \jump -  \frac{\jump^\dagger \jump}{2} +  \frac{\braket{\jump^\dagger \jump}}{2} +  \frac{\beta \braket{\jump^\dagger+ \jump}}{2} \right)\right]\ket{\psi(t)},
\end{split}
\end{equation}

The ideal limit of homodyne detection is when the coherent field amplitude goes to infinity.
We stress that, in this limit, the number of detections per time unit is infinite, and a stochastic Schr\"odinger equation based on $\de N(t)$ is ill defined.
Thus, some care needs to be taken in order to derive the appropriate limit $\beta\to \infty$.
From a physical point of view, the condition $\beta\gg 1$ implies that quantum jumps will occur more frequently, but at the same time their effect on the evolution of $\ket{\psi(t)}$ is smaller.
Indeed, the detected field is almost entirely due to the coherent field, and thus the measure backaction on the system must be extremely small.
Here, we report only the final result (the formal derivation is provided in, e.g., Ref.~\cite{Wiseman_BOOK_Quantum}):
\begin{equation}
\label{Eq:Diffusive_Homodyne}
\begin{split}
\de \ket{\psi(t)}& = \left[ \de W(t)   \left(\jump - \frac{\braket{\jump^\dagger + \jump}}{2} \right) \right.
\\ &  \left.  \qquad+ \tau \left(-\ii \hat{H} - \frac{\jump^\dagger \jump}{2} +\jump \frac{\braket{\jump^\dagger + \jump}}{2} - \frac{\braket{\jump^\dagger+\jump}^2}{8} \right)\right] \ket{\psi (t)},
\end{split}
\end{equation}
where $\de W(t)$ is a Wiener process of variance $\tau$ and mean 0 \cite{Gardiner_BOOK_Quantum,Gardiner_BOOK_Stochastic}.
We call a solution of this equation a \emph{homodyne quantum trajectory}.

\subsection{Physical interpretation of a quantum trajectory}
\label{Sec:Physical_interpretation}
\begin{figure}
\centering
\includegraphics[width=0.7\linewidth]{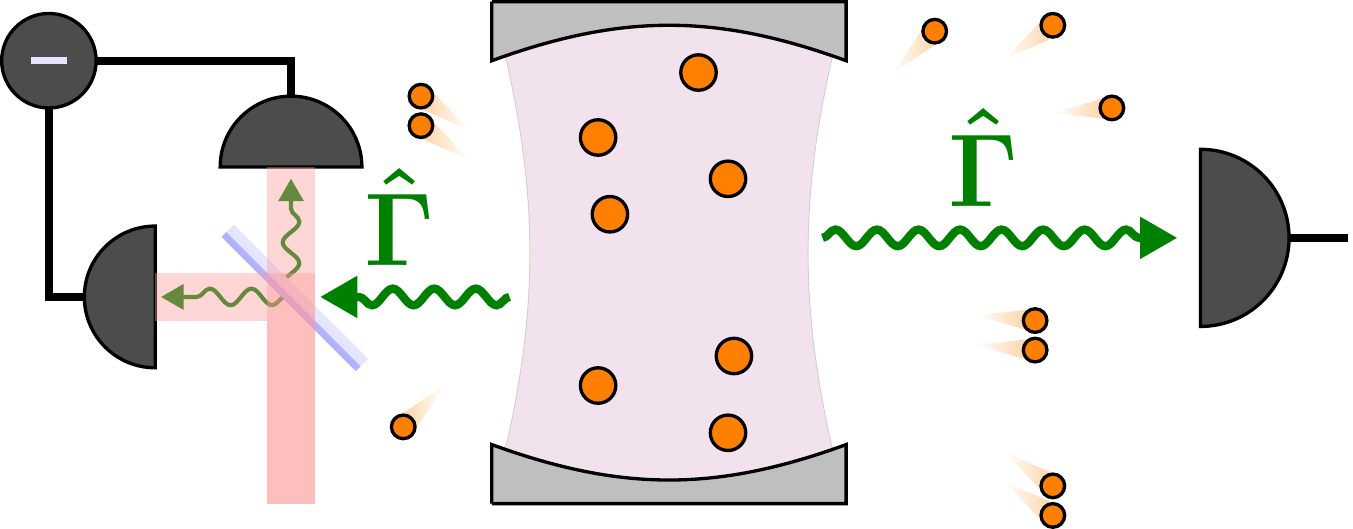}
\caption{Sketch of the two schemes of detection on a photonic cavity. On the right side, the photon counting mechanism.
Every time a perfect photodetector ``clicks'', an observer knows that a photon has been lost by the cavity.
Homodyne detection on the left.
Before the measure by a perfect photodetector, the output filed of the cavity is mixed (e.g., with a beamsplitter) with a strong local filed.
The statistics of ``clicks'' allows an observer to follow the state of the cavity.
}
\label{fig:Homodine_vs_photonocountung}
\end{figure}

A natural question is what is the relation between \eqref{Eq:SSE} and \eqref{Eq:Lindblad_Master_Equation}.
In agreement with the previous discussion, one can think of the Linblad master equation as a continuous unread measure performed on the system, while in \eqref{Eq:SSE} one keeps track of the measure results.
In the same way in which the mean over an infinite number of read measures must coincide with the expectation value of an unread one, it is possible to recover the result of the Lindblad master equation by averaging over an infinite number of quantum trajectories  \cite{Gardiner_BOOK_Quantum,Carmichael_BOOK_2,Haroche_BOOK_Quantum,DaleyAdvancesinPhysics2014}.

Furthermore, one may try to provide some meaning to individual trajectories.
Indeed, a single quantum trajectory corresponds to the simulation of an ideal experiment, in which the environment is continuously monitored by perfect instruments.
In this regard, single quantum trajectories can account for observed features.
However, the result obtained along a single quantum trajectory can strongly depend upon the choice of Kraus operators, and some care should be taken about which properties can be inferred.

\subsubsection{Photon counting vs Homodyne}

Consider, for instance, the counting trajectory derived in Sec.~\ref{Sec:Conunting_Trajectories} in the specific case of an optical cavity, where $\hat{\Gamma} = \sqrt{\gamma} \hat{a}$ (Fig.~\ref{fig:Homodine_vs_photonocountung}).
Indeed, suppose we are able to build a perfect photodetector which continuously \emph{measures the environment} and ``clicks'' every time it registers a photon.
If we hear a ``click'', we know for sure that a jump has occurred and the cavity wave function has undergone a quantum jump. 
If no jump has occurred, instead, the system evolves under the effective Hamiltonian $\hat{H}_{\rm eff}=\hat{H} -\ii  \gamma \hat{a}^\dagger \hat{a} /2$.\footnote{In this discussion, we will neglect the terms ensuring the normalization of the wave function.}
Remarkably, the absence of a quantum jump does not mean that the system evolves under the effect of the Hamiltonian alone, but knowing that a jump has not occurred gives us information about the state of the cavity.
Indeed, the imaginary term $-\ii  \gamma \hat{a}^\dagger \hat{a}$ is the backaction of the continuous measurement of the photodetector.

Consider now \eqref{Eq:SSE_Homodyne} in the limit $\beta\gg 1$.
Since the cavity output field is mixed with a local reference oscillator, the detector measures a superposition of the two fields.
Since the local laser is extremely strong, the detector continuously ``clicks'', and the field inside the cavity undergoes a quantum jump.
However, as it stems from \eqref{Eq:SSE_Homodyne}, the effect of such a quantum jump is minimal:
\begin{equation}
\frac{\jump+\beta}{\sqrt{\braket{(\jump^\dagger +\beta) (\jump+\beta)}}}  - \mathds{1} = \frac{\jump}{\beta} - \frac{\braket{\jump +\jump^\dagger}}{2 \beta} + \mathcal{O} (\beta^{-2}).
\end{equation}
Indeed, the detection will be almost entirely due to the local oscillator, and therefore the cavity wavefunction remains almost unaffected by the measure. 

From an information theory perspective, the difference between the two types of trajectories reduces to the way an observer acquires information about the state of the cavity.
In the counting case, the state is abruptly and randomly modified due to the great amount of information gained by one detection.
In the homodyne case, instead, the state is continuously randomly changed due to the high number of detections. 
However, the information gained about the state of the cavity is minimal since every registered photons is in a superposition of the cavity output field and of the local oscillator.

\begin{figure}
\centering
\includegraphics[width=0.98\linewidth]{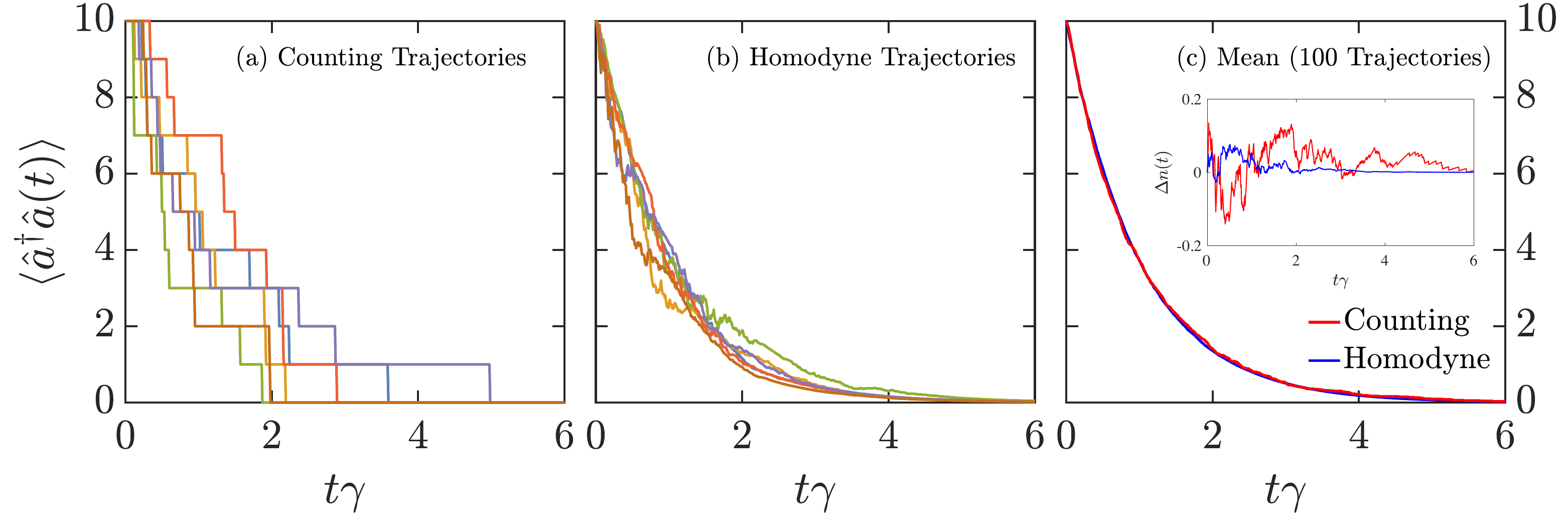}
\caption{Mean number of photon $\braket{\hat{a}^\dagger \hat{a} (t)}$ as a function of time for a resonator subject to dissipation.
	Panel (a): Counting trajectories. 
	The wave function changes abruptly under the effect of a quantum jump.
	Panel(b): Homodyne trajectories.
	The evolution of the wave function is a noisy one.
	Panel (c): Average $\braket{\hat{a}^\dagger \hat{a} (t)}$  over 100 trajectories.
	The two procedures recover the same results.
	Inset: $\Delta n (t) = n(t) - \braket{\hat{a}^\dagger \hat{a} (t)}$ as a function of time, where $n(t)$ is the mean number of photons obtained via direct integration of the master equation with a cutoff of 30 photons.
	Parameters: $\omega/\gamma=1$.
	}
\label{fig:Homodine_vs_photoncounting_trajectory}
\end{figure}

To visualize these differences in the procedures, we simulate the radiative damping described in Sec.~\ref{subsec:LindbladForm} and Sec.~\ref{Sec:Quantum_jumps}.
We consider a resonator, described by an Hamiltonian $\hat{H}= \omega \hat{a}^\dagger \hat{a}$, $\omega$ being the energy of one photon in the cavity and $\hat{a}$ the annihilation operator. The coupling to the environment is via the single jump operator $\jump=\sqrt{\gamma}\hat{a}$, where $\gamma$ is the mean-lifetime of one photon inside the resonator. At $t=0$, we initialize the cavity is the Fock state with ten photons, i.e. $\ket{\psi (t=0)}=\ket{n=10}$.
In Fig.~\ref{fig:Homodine_vs_photoncounting_trajectory} we plot the mean number of photons $\braket{\hat{a}^\dagger \hat{a}(t)}$ along five counting and five homodyne quantum trajectories.
In the case of a counting trajectory [panel (a)], the evolution of the parameter $\braket{\hat{a}^\dagger \hat{a}(t)}$ is smooth, except in a finite number of points, where a quantum jump happens.
As for the homodyne [panel (b)], instead, there are no abrupt jumps, but the evolution is a noisy one.
Both procedures, once the average is taken, recover the same result [panel (c)], which coincides with the one obtained via integration of  the Lindblad master equation [inset of panel (c)].

\subsubsection{Not detecting is still measuring}
\label{Sec:Dissipative_State_transfer}

We show here an example that can clarify the physics of open quantum systems continuously monitored.

Consider a single qubit, with Hamiltonian $\hat{H} = \frac{\omega_1}{2} \hat{\sigma }_z$, and dissipating via $\gamma \DD[\hat{\sigma}_{-}] $.
The effective Hamiltonian reads
$\hat{H}_{\rm eff}= \omega/2 \hat{\sigma}_z - i \gamma/2 \, \hat{\sigma}_+ \hat{\sigma}_-$.
Assume that system is initialized in the superposition state $\ket{\psi_0} = (\ket{e}+\ket{g})/{\sqrt{2}}$.
While the state $\ket{g}$ does not decay under the action of the effective Hamiltonian, $\ket{e}$ does.

The environment back-action through the continuous measure is clear, as described in Refs.~\cite{Molmer1993}:
\textit{ each time there is no quantum jump} [$\de N(t) =0$] \textit{the weight of the state} $\ket{g}$ \textit{increases}.
Equivalently, the observer gains information on the wave function, and the fact that there was no jump makes the state $\ket{g}$ more probable \cite{Haroche_BOOK_Quantum}.
Thus, the non-Hermitian part of \eqref{Eq:Effective_Hamiltonian_counting} continuously affects the wave function, which approaches the ground state as time passes and no detection occurs. 
This can be seen as a Bayesian update of the wave function \cite{ChantasriPRX16}.

Consider now two noninteracting dissipative qubits.
Their Hamiltonian reads:
\begin{equation}
\label{Eq:Hamiltonian}
    \hat{H} = \frac{\omega_1}{2} \hat{\sigma }_z^{(1)}+ \frac{\omega_2}{2} \hat{\sigma}_z^{(2)}.
\end{equation}
where $\hat{\sigma}_z^{(j)}$ is the Pauli $z$ operator of the $j$-th qubit.
We indicate a state as, e.g., $\ket{e,g}$, meaning that qubit 1 is in the excited state and qubit 2 is in the ground state. 
We assume that each qubit dissipates its excitation into the environment both {\emph {locally}} and {\emph {collectively}}, using the terminology of Ref.~\cite{ShammahPRA18}.
Thus, the dissipators read:
\begin{equation}\label{Eq:jumps}
 \gamma_1 \DD\left[\hat{\sigma}_{-}^{(1)} \right]   ,
\quad    \gamma_2 \DD\left[\hat{\sigma}_{-}^{(2)} \right] ,
\quad     \gamma_c \DD\left[\frac{\hat{\sigma}_{-}^{(1)}+ \hat{\sigma}_{-}^{(2)}}{\sqrt{2}}\right],
\end{equation}
where $\hat{\sigma}_\pm^{(j)}=(\hat{\sigma}_x^{(j)} \pm i \hat{\sigma}_y^{(j)})/2$ is the lowering (raising) operator of the $j$-th qubit.
While the first two dissipators characterize the dynamics of noninteracting qubits, the collective dissipation and its rate $\gamma_c$ naturally emerges when the qubits are sufficiently close to each other with respect to the typical wavelength of the electromagnetic reservoir (see the discussion in, e.g., \cite{MacriPRA20}). 
Furthermore, the collective dissipation can be engineered for noninteracting nonlinear cavities, as discussed in Ref.~\cite{MamaevQuantum18}, such as circuit-QED systems \cite{Gu2017}.
For strong (i.e., infinite) photon-photon interaction, the physics of the nonlinear resonators becomes that of a two-level system  \cite{CarusottoRMP13,CarusottoPRB2005,BirnbaumNat05,DelteilNatMat19,LangPRL11}.
We will suppose hereafter to initialize the qubit always in the state $\ket{e,g}$, i.e., the first qubit is excited, and the second one is in its ground state.

\begin{figure}
    \centering
    \includegraphics[width=0.4 \textwidth]{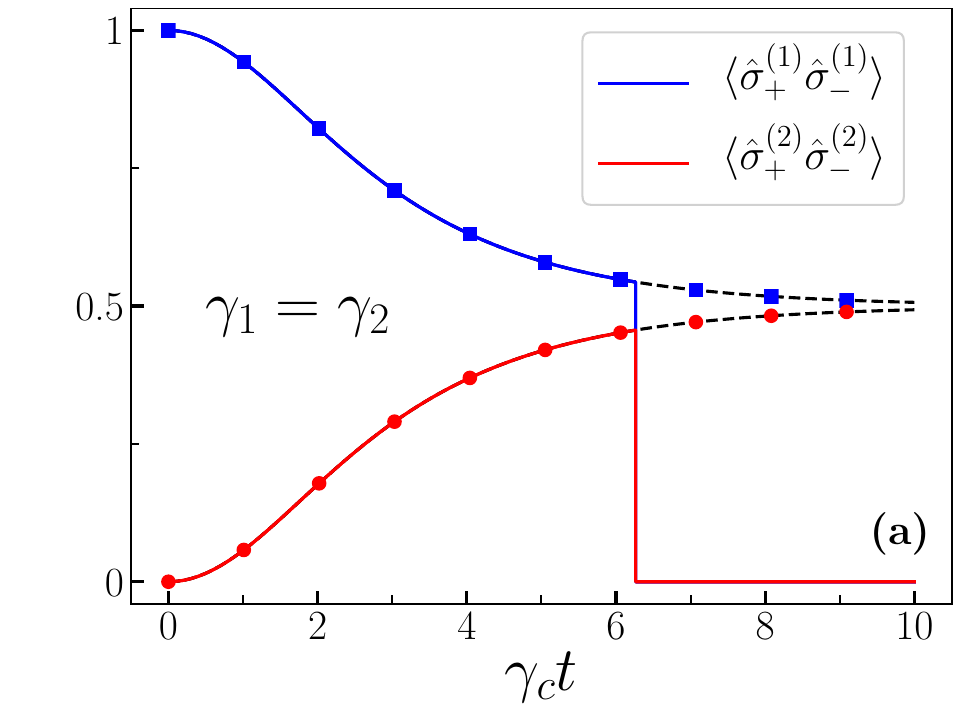}
    \includegraphics[width=0.4 \textwidth]{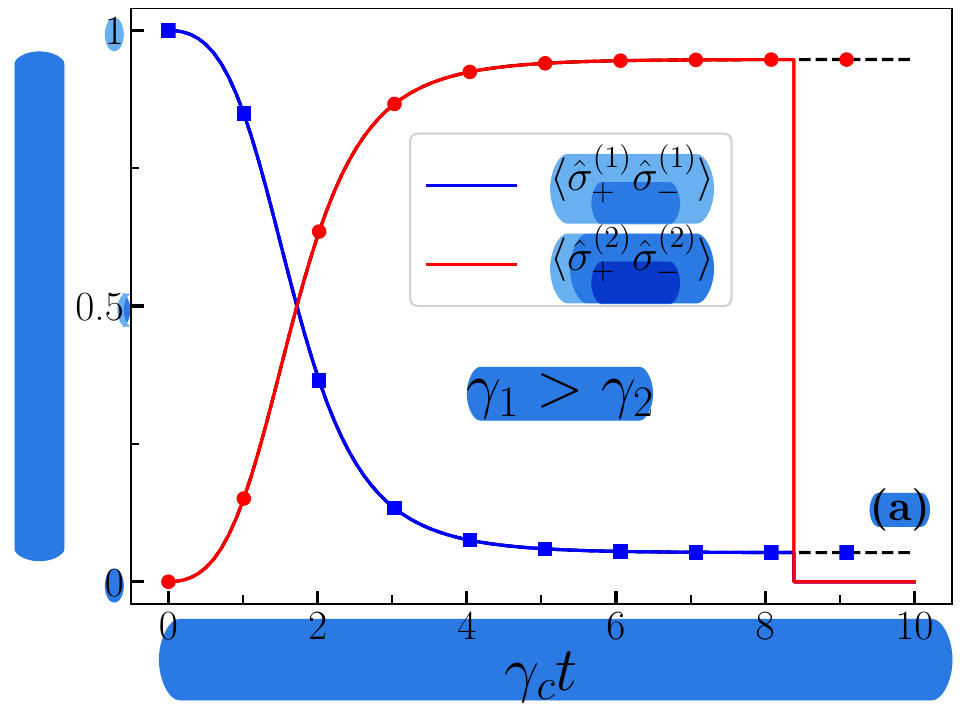}
    \caption{As a function of time, $\hat{\sigma}_+^{(1)}\hat{\sigma}_-^{(1)}$ (qubit 1, blue curve and squares, starting from 1) and $\hat{\sigma}_+^{(2)}\hat{\sigma}_-^{(2)}$ (qubit 2, red curve and circles, starting from 0) for an ideal counting quantum trajectory. Left: dynamics in the presence of collective dissipation and local dissipation identical for qubits 1 and 2. As time passes, the wave function tends towards the Bell state $\ket{\Psi^-}$, until a quantum jump occurs at time $\gamma_c t \simeq 6$. 
    The black dashed lines represent the dynamics without quantum jumps. 
    Right: dynamics in the presence of collective dissipation and unbalanced local dissipation. While collective disipation would create a Bell state $\ket{\Psi^-}$, the unbalance in local dissipation transforms $\ket{\Psi^-}$ into a state approximately $\ket{g,e}$. These two processes ensure the excitation swap until a quantum jump occurs at $\gamma_c t \simeq 8$. Images modified from Ref.~\cite{MingantiPRA21}.}
    \label{fig:transfer}
\end{figure}

If $\gamma_c =0 $, since the qubits are non-interacting the only thing that happens is that, sooner or later, a photon is emitted by the first qubit.
If, instead, $\gamma_1=\gamma_2$, but $\gamma_c \neq 0$, we observe that also the second qubit gets excited as time passes, and if no jump as occurred (see left panel of Fig.~\ref{fig:transfer}).
Finally, if we are to increase the dissipation rate of the first qubit ($\gamma_1 > \gamma_2$), the effect becomes even more counterintuitive: if at long time no emission occurred the state ``passes'' in the second (originally unexcited) qubit. How is this possible?

In the absence of quantum jumps the evolution of the system is dictated solely by the effective Hamiltonian.
Thus, no quantum jumps (i.e., ``no clicks'' of the detectors) affect the system wave function via the renormalization induced by the non-Hermitian terms of $\hat{H}_{\rm eff}$ \cite{Molmer1993}.
Generally speaking, given a superposition of ``bright'' and ``dark'' states of the dissipation, each time there is no quantum jump the wave function becomes more populated by the dark state.

To show that this is indeed the case, let us now consider $\gamma_1=\gamma_2$.
In this case, there exist two dark states, according to the number of excitations in the system.
Indeed, by introducing the Bell states $\ket{\Psi^\pm}=(\ket{e, g} \pm \ket{g, e})/\sqrt{2}$, we have 
\begin{equation}
    \ket{e,g}=\left( \ket{\Psi^+} + \ket{\Psi^-} \right)/\sqrt{2}.
\end{equation}
Since we always initialize the system in $\ket{e,g}$, the collective dissipator reads
\begin{equation}\label{Eq:collective_dissipator_Bell}
   \gamma_c \mathcal{D}\left[\frac{\hat{\sigma}_{-}^{(1)}+ \hat{\sigma}_{-}^{(2)}}{\sqrt{2}}\right] \rhot = \gamma_c \mathcal{D}\left[\ket{g,g}\bra{\Psi^+}\right] \rhot.
\end{equation}
Thus, both $\ket{g,g}$ and $\ket{\Psi^{-}}$ are dark states of the dissipator in \eqref{Eq:collective_dissipator_Bell}, while $\ket{\Psi^{+}}$ is the bright state.

The dynamics of the initial state $\ket{e,g}$ is, therefore, characterized by two possible processes.
If a quantum jump occurs, the wave function at time $t$ becomes immediately $\ket{g,g}$ and there is no more dynamics.
Instead, when no quantum jump happens, the state $\ket{\Psi^-}$ becomes more probable as time passes.
This process is captured by projecting the effective Hamiltonian in \eqref{Eq:Effective_Hamiltonian_counting} onto the one-excitation manifold, which in the presence of only the collective dissipation (up to a constant) reads 
\begin{equation}
\sum_{j = \pm}
   \expect{\Psi^{j}}{\hat{H}_{\rm eff}}{\Psi^{j}} \ket{\Psi^{j}}\bra{\Psi^{j}} = - i \frac{\gamma_c}{2} \ket{\Psi^{+}}\bra{\Psi^{+}}.
\end{equation}
Thus, when no quantum jump happens the coefficient of the state $\ket{\Psi^{+}}$ exponentially decays to zero, and therefore $\ket{\Psi(t)}$ approaches $\ket{\Psi^-}$.
We show this effect in Fig.~\ref{fig:transfer} (left), where $\ket{e, g}$ converges towards $\ket{\Psi^{-}}$ before a quantum jump takes place at $\gamma_c t \simeq 6$.
The red and blue curves represent the expectation value of $\hat{\sigma}^{(1,\,2)}_+\hat{\sigma}_-^{(1,\,2)}$ obtained from a quantum trajectory simulation.

A similar calculation can be performed in the configuration $\gamma_1>\gamma_2$ and $\gamma_c =0$, which is characterized by a different set of bright and dark states, leading to the dynamics in Fig.~\ref{fig:transfer} (right).
For more details, see Ref.~\cite{MingantiPRA21}.

\section{Methods for many-body systems}
\label{Chap:Many_body}

The physics of open quantum systems is further enriched when the behaviour of many interacting bodies is considered.
This additional layer of complexity calls for methods able to simulate the dynamcs and/or the stationary state of the system under consideration \cite{Orus_review}.

In what follows we will briefly describe two classes of methods suitable for matrix product operators (MPOs) \cite{BiellaPRA15,Lars2014,ZwolakPRL04,VestraetePRL04} and cluster mean-field (CMF) \cite{JinPRX16,JinPRB18,BiellaPRB2018,BiellaPRA17} ansatzes. These have been shown to be effective and reliable to simulate $1$ and $2D$ lattice systems, respectively. 

\subsection{Matrix-product-operator approach}
For the time evolution toward the steady state of the system, it is possible to exploit a method based on the time-evolving block decimation (TEBD) \cite{VidalPRL04,VidalPRL03} scheme extended to open systems. In this framework the density matrix for a chain of $M$ sites with open boundary conditions
\be
\rho = \sum_{i_\alpha, j_\alpha=1}^d \ R_{i_1 \cdots i_L, \, j_1 \cdots j_M} \, || i_1 \cdots i_M , \, j_1 \cdots j_M \rangle \rangle
\ee
is written as a matrix-product-state (MPS) in the enlarged Hilbert space of dimension $d^2$, 
where $d$ is the dimension of the local Hilbert space $\mathbb{H}$.
The repeated application of singular value decompositions 
of the tensor $R$ 
leads to the following representation:
\be
\label{MPO}
\rho = \sum_{i_\alpha, j_\alpha=1}^d \sum_{\alpha,\beta,\dots,\gamma=1}^{\chi} \ B_{1,\alpha}^{[1]i_1,j_1}\lambda_{\alpha}^{[1]} B_{\alpha,\beta}^{[2]i_2,j_2}\lambda_{\beta}^{[2]}\dots  \lambda_{\gamma}^{[M-1]}B_{\gamma,1}^{[M]i_M,j_M}  || i_1 \cdots i_M , \, j_1 \cdots j_M \rangle \rangle .
\ee
Here $|| i_1 \cdots i_M , \, j_1 \cdots j_M \rangle \rangle = \bigotimes_{a=1}^{M}\ket{i_a}\bra{j_a}$ 
represents a basis for the density matrix in the product Hilbert space 
$\mathbb{H}^{\otimes M}\otimes\mathbb{H}^{\otimes M}$. 

The time evolution of the MPO \eqref{MPO} is performed rewriting the evolution operator via Trotter decomposition as 
\begin{equation}
\label{trotter}
    e^{-i dt \supmat{\LL}} \simeq 
    \left(\prod_{j=1}^{M-1} e^{-i  \mathsf{L}^{j,j+1} dt/2}\right)
    \left(\prod_{j=1}^{M-1} e^{-i  \mathsf{L}^{M-j,M-j+1} dt/2}\right) + \mathcal{O}\left(dt^3\right),
\end{equation}
where $dt$ is the time step and $\supmat{\LL}=\sum_{j=1}^{M-1}\mathsf{L}^{j,j+1}$ is a generic Lindbladian of a system with local and nearest-neighbour interactions in the written in its matrix form (see App.\ref{sec:LiouvMatr}) according to the prescription used for the vectorization of the density matrix.
After each application of local two-site gates in \eqref{trotter} on the state \eqref{MPO} the bond link of the MPO representation increases as $\chi\to\chi d^2$. 
The evolution can be thus performed {\itshape exactly} until $\chi d^2<\chi_{\rm max}$ where $\chi_{\rm max}$ is the maximum bond dimension allowed in the simulation.
When this condition is violated one must trunctate the representation keeping only the $\chi_{\rm max}$ largest singular values $\lambda_{\alpha}^{[i]}$.
This choice is well justified the Schmidt spectrum $\lambda_{\alpha}^{[i]}$ decays fast enough and thus depends on the specific region of the parameters under consideration.
In practice one has to check if the observables under investigation
converge as $\chi_{\rm max}$ is increased for the time range considered in the simulation.

This scheme known as TEBD is repeated until the steady is reached.
The amount of correlation generated in the dynamics does not increase indefinitely but saturates as the stationary regime is approached.
This leads to the possibility of reaching dynamically the steady state of a system within this framework with finite computational resources.

\subsubsection{Stochastic trajectories with Monte Carlo matrix product states}

Matrix product states can be also used for the simulation of stochastic quantum trajectories \cite{DaleyAdvancesinPhysics2014,GemmelmarkPRA2010,MakiSciPost2023,VovkPRL2022,PreisserPRA2023}.
According to the quantum jump protocol, after each time step $dt$, we stochastically choose whether to evolve the system with the non-Hermitian effective Hamiltonian or to perform a quantum jump. For simplicity here we will restrict ourselves to the case of single-site jump operators. 

\begin{figure}
    \centering
    \includegraphics[scale = 0.18]{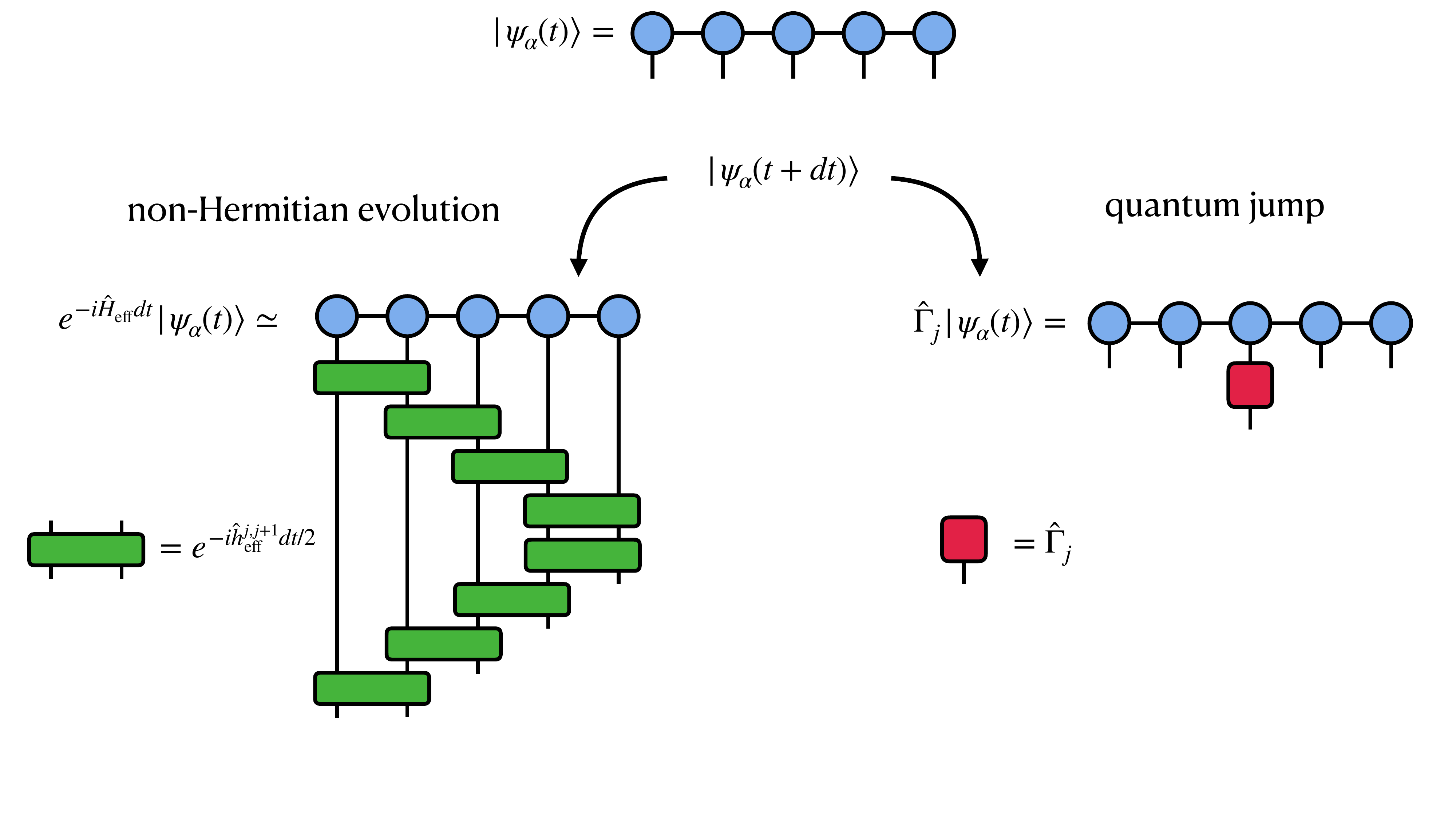}
    \caption{A sketch of the MCMPS method. 
    With probability $1-\sum_{i=1}^L p_i(t)$ and, the system  evolves according to the effective non-Hermitian Hamiltonian $H_{\rm eff}$ in its Trotterized form \eqref{trotter} (left side) or, with probability $p_j(t)$, undergoes a quantum jump. In this a quantum jump occurs on the $j$-th site (right side).}
    \label{fig:MPS_QT}
\end{figure}

The first case occurs with probability 
$1-\sum_{i=1}^L p_i(t)$ where $p_i(t)=dt \braket{\hat{\Gamma}^\dagger_i\hat{\Gamma}_i}_t$ (where $\{\hat{\Gamma}_i\}$ is the set of jump operators) and $\braket{\bullet}_t$ is the expectation value computed along a single trajectory.
Accordingly, we evolve the $\alpha$-th trajectory as 
    \begin{equation}
    \label{Heffevo}
        |\psi_\alpha(t+dt)\rangle = \frac{e^{-i H_{\rm eff} dt} |\psi_\alpha(t)\rangle}{\lVert e^{-i H_{\rm eff} dt} |\psi_\alpha(t)\rangle \rVert},
    \end{equation}
having defined
\be
\hat H_{\rm eff} = \hat{H}-\frac{i}{2}\sum_i\hat{\Gamma}^\dagger_i\hat{\Gamma}_i.
\ee
The evolution \ref{Heffevo} can be performed via standard TEBD scheme upon writing the trajectory state as a MPS (here we again assume $\hat H_{\rm eff}=\sum_{i=1}^{M-1}\hat{h}_{\rm eff}^{i,i+1}$). 
Otherwise, with probability $p_i(t)$ we apply the $i$-th jump operator:
    \begin{equation}
        |\psi_\alpha(t+dt)\rangle = \frac{\hat{\Gamma}_i|\psi_\alpha(t)\rangle}{\lVert \hat{\Gamma}_i|\psi_\alpha(t)\rangle\rVert}.
    \end{equation}
The action of a given quantum jump can be easily computed by applying the local operator
$\hat{\Gamma}_i$ to the MPS structure.
The whole procedure goes under the name of Monte Carlo matrix product state (MCMPS) and is illustrated in Fig.~(\ref{fig:MPS_QT}). 

\subsection{Cluster mean field approach}
Here we describe the cluster mean-field (CMF) approach that allow for the effective simulation of the dynamics in $D\geq 2$ open (hyper)lattices.
For simplicity we will now focus on the case of Hamiltonian and dissipators containing local and nearest-neighbour interactions only 
\be
\hat H = \sum_i \hat h_i +\sum_{\braket{i,j}} \hat h_{ij}, \qquad
\mathcal{L}[\rho] = \sum_i\mathcal{L}_i[\rho] + \sum_{<i,j>}\mathcal{L}_{ij}[\rho],
\ee
with
\be
\mathcal{L}_i[\rho] = \gamma_1\left(\hat{L}_i\rho\hat{L}_i^\dagger - \{\hat{L}_i^\dagger\hat{L}_i;\rho\}/2\right), \ \ 
\mathcal{L}_{ij}[\rho] = \gamma_2 \left(\hat{M}_{ij}\rho\hat{M}_{ij}^\dagger - \{\hat{M}_{ij}^\dagger\hat{M}_{ij};\rho\}/2\right),
\ee
with $\hat{M}_{ij}=\hat{l}_i\hat{m}_j$ accounting for one- and two-site jump operators at a rate proportional to $\gamma_{1,2}$, respectively.
We now propose a cluster Gutzwiller ansatz for the system density matrix 
\be
\rho(t) = \bigotimes_{\mathcal{C}} \rho_\mathcal{C}(t),
\ee
where $\rho_\mathcal{C}$ is a local density matrix with support over the cluster $\mathcal{C}$ of contiguous sites. 
In a CMF approach the dynamics inside a cluster $\mathcal{C}$ is described by a CMF Hamiltonian 
\be
\hat{H}_{\rm CMF}(t)=\hat{H}_{\mathcal{C}} +  \hat{H}_{\mathcal{B}(\mathcal{C})}(t),
\ee
where $\hat{H}_{\mathcal{C}}=\sum_{i\in\mathcal{C}}\hat h_i + \sum_{\braket{i,j}|i,j\in\mathcal{C}}\hat h_{ij}$ describes the local interactions inside the cluster and the time-dependent boundary Hamiltonian
\be
\hat{H}_{\mathcal{B}(\mathcal{C})}(t) = \sum_{j\in{\mathcal{B}(\mathcal{C})}}
\vec{B}_j(t)\cdot\hat{\vec{\sigma}}_j
\ee
accounts for the mean-field interaction of the cluster $\mathcal{C}$ with its neighbours $\mathcal{C'}$ where $\hat{\vec{\sigma}}_j=[\ssx_j,\ssy_j,\ssz_j]$ and $\vec{B}_j(t)=[B^x_j(t),B^y_j(t),B^z_j(t)]$ are time-dependent mean-field parameters to be determined self-consistently as $\vec{B}_j (t)= f\left({\rm Tr}\left[\rho_{\mathcal{C'}} (t)\hat{\vec{\sigma}}_i\right]\right)$, $i\in\mathcal{C'}$ being the proper neighbour site of $j\in\mathcal{C}$ and $f(\bullet)$ being a linear function of the magnetization determined by the specific nearest-neighbour interaction.
In its most general case the nearest-neighbour interaction takes the form $\hat h_{ij} = \sum_{\alpha,\beta=x,y,z} J_{\alpha\beta} \ \hat\sigma_i^\alpha \hat\sigma_j^\beta$ and thus for the mean-field parameters we get
\be 
\qquad {B}^{\beta=x,y,z}_j(t)= \sum_{\alpha=x,y,z} J_{\alpha\beta}{\rm Tr}\left[\rho_{\mathcal{C'}} (t)\hat{\sigma}^\alpha_i\right].
\ee

For the dissipative part of the dynamics we can do the same kind of reasoning. We use the decomposition
\be
\mathcal{L}_{\rm CMF}(t)[\rho_{\mathcal{C}}] = \mathcal{L}_\mathcal{C}[\rho_{\mathcal{C}}] + \mathcal{L}_{\mathcal{B}(\mathcal{C})}(t)[\rho_{\mathcal{C}}],
\ee
where 
$
\mathcal{L}_\mathcal{C}[\rho_{\mathcal{C}}] = \sum_{i\in\mathcal{C}} \mathcal{L}_i[\rho_{\mathcal{C}}]
+\sum_{\braket{i,j}\in\mathcal{C}} \mathcal{L}_{ij}[\rho_{\mathcal{C}}]
$
accounts for the incoherent processes inside the cluster and  
\be
\label{boundarydiss}
\mathcal{L}_{\mathcal{B}(\mathcal{C})} [\rho_{\mathcal{C}}]= \sum_{j\in{\mathcal{B}(\mathcal{C})}} \Gamma_{j}(t)\left(\hat{m}_j\rho_{\mathcal{C}}(t)\hat{m}_j^\dagger -\frac12 \left\{\hat{m}_j^\dagger\hat{m}_j,\rho_{\mathcal{C}}(t)\right\}\right),
\ee
for incoherent events connecting the cluster $\mathcal{C}$ with the neighbours.
Here we defined the mean-field time-dependent parameters $\Gamma_j(t)$ that also requires to be computed self-consistently.
For the model we are considering in this paper we get
\be
\Gamma_j(t)=\gamma_2 \ {\rm Tr}\left[\rho_{\mathcal{C'}}(t) \hat{l}_i^\dagger\hat{l}_i\right].
\ee
In practice the Gutzwiller decoupling renormalizes the bare two-spin dissipation rate $\gamma_2$ by a factor that depends on the state of the neighbour cluster $\mathcal{C'}$.
Finally, if we require translational invariance with respect to the cluster periodicity $\rho_{\mathcal{C}}=\rho_{\mathcal{C'}}, \forall\mathcal{C},\mathcal{C'}$ we find a closed effective muster equation of the form 
\be
\frac{d \rho_{\mathcal{C}} }{dt} =  -\frac{i}{\hbar}\left[\hat{H}_{\rm CMF}(t), \rho_{\mathcal{C}}\right] 
+ \mathcal{L}_{\rm CMF}(t)[\rho_{\mathcal{C}}].
\ee

\textbf{Acknowledgements}
\begin{acknowledgement}
We acknowledge the Institute Pascal (Université Paris-Saclay, Orsay, France) for hosting us in the context of the workshop “Open QMBP 2023”.
We are grateful to the organizers of the event together with the administrative support for providing us with a stimulating environment during the program. 
The authors are indebted to the many colleagues working in this field for discussions, suggestions, and comments over the years. 
AB acknowledges fundings by the European Union - NextGeneration EU, within PRIN 2022, PNRR M4C2, Project TANQU 2022FLSPAJ [CUP B53D23005130006].
\end{acknowledgement}

\appendix

\section{Radiative damping in an optical cavity: three different derivations}

All through the manuscript, we assumed either the form of the Kraus operators and the corresponding measurements, or the corresponding jump operators.
These approaches do not tell much about the operators $\hat{\Gamma}_\mu$.
Here, we discuss how to derive the radiative damping of an optical cavity through  a microscopic description, using measurement theory, and assuming the presence of quantum jumps in a quantum trajectory.

\subsection{Microscopic derivation}

Here we derive the evolution equation starting from a microscopic description of the electromagnetic field in an optical cavity.

Let us consider a one-mode optical cavity, where the EM field is confined between two high-quality mirrors.
In first approximation, all photons inside the cavity have the same frequency $\omega_c$.
Thus, discarding the constant $\omega/2$ term, the Hamiltonian reads
\begin{equation}
\label{Eq:Cavity_Alone}
\hat{H}_C= \omega_c \hat{a}^\dagger \hat{a}.
\end{equation}
The environment, instead, is described as the collection of infinitely many harmonic oscillators.
Therefore, its Hamiltonian is
\begin{equation}
\hat{H}_E= \int_0^{\infty} \de \omega \, \omega \, \hat{b}^\dagger(\omega) \hat{b}(\omega).
\end{equation}
We suppose that the system and the environment are coupled via
\begin{equation}
\hat{H}_I= \int \de \omega \, g(\omega) (\hat{a} + \hat{a}^\dagger) (\hat{b}(\omega)+\hat{b}^\dagger(\omega)).
\end{equation}
The evolution of the cavity coupled to the environment is
\begin{equation}
-\ii \partial_t \ket{\Psi} = (\hat{H}_C + \hat{H}_E + \hat{H}_I) \ket{\Psi},
\end{equation}
where $\ket{\Psi}$ is the wave function describing the system and the environment.

To simplify the problem, we pass in the interaction picture, i.e. we introduce $\ket{\tilde{\Psi}} = \hat{U}(t) \ket{\Psi}$, where $\hat{U}(t)=\exp[\ii (\hat{H}_C + \hat{H}_E)  t]$.
We have
\begin{equation}
-\ii \partial_t \ket{\tilde{\Psi}}= \hat{\tilde{H}} \ket{\tilde{\Psi}},
\end{equation}
where the interaction Hamiltonian
\begin{equation}
\hat{\tilde{H}} =  \hat{U}(t) \hat{H}_I  \hat{U}^\dagger(t) = \int_0^{\infty} \de \omega g(\omega) \left(\hat{a} e^{-\ii \omega_c t} + \hat{a}^\dagger e^{\ii \omega_c t}\right) \left( \hat{b}(\omega) e^{-\ii \omega t} + \hat{b}^\dagger(\omega) e^{\ii \omega t} \right)
\end{equation}
can be easily obtained exploiting $e^{\ii \omega_c \hat{a}^\dagger \hat{a} t} \hat{a} = e^{-\ii \omega_c t} \hat{a}e^{\ii \omega_c \hat{a}^\dagger \hat{a} t}$ (a similar relation exists for $\hat{b}(\omega)$).
When we expand the product of exponentials, we get two terms: one depending on $\omega_c - \omega$ and one on $\omega_c +\omega$.
We can thus perform a secular approximation, that is, in the limit of small interactions we can neglect the fast oscillating terms.
Intuitively, when integrating over a period $t= 2\pi/(\omega_c - \omega)$, the fast frequencies will average to zero in the Schr\"odinger equation.
This simplification, known as rotating wave approximation, gives
\begin{equation}
\hat{\tilde{H}}(t) =  \int_0 ^\infty \de \, \omega g(\omega) \left(\hat{a} \hat{b}^\dagger(\omega) e^{-\ii (\omega_c-\omega) t} + \hat{a}^\dagger \hat{b}(\omega) e^{\ii (\omega_c - \omega) t}\right) .
\end{equation}
Again, if $g(\omega)$ is small, in the previous integral the only terms which will be relevant are those for which $\omega_c-\omega \simeq 0$.
Thus, we can send the integration limit towards $-\infty$.
Moreover, we suppose a sufficiently regular $g(\omega)$, so that $g(\omega) \simeq g(\omega_c) \equiv \sqrt{\gamma/2\pi}$ if $\omega \simeq \omega_c$.
Physically, $\gamma$ represent the decay rate (i.e. the inverse of the lifetime) of a photon inside the cavity.
Hence
\begin{equation}
\hat{\tilde{H}}(t) =  \sqrt{\frac{\gamma}{2 \pi}} \left(\hat{a} \hat{b}^\dagger(t) + \hat{a}^\dagger \hat{b}(t) \right),
\end{equation}
where we define
\begin{equation}
\hat{b}(t)\equiv \int_{-\infty}^{+\infty} \de \omega \, \hat{b}(\omega) e^{\ii (\omega_c - \omega) t}.
\end{equation}
The definition of $\hat{b}(t)$ implies that $[\hat{b}(t), \hat{b}^\dagger(t')]=\delta(t-t')$.
Indeed, we are assuming that the environment does not have memory about its previous states, i.e. it is \emph{Markovian}.

We are now interested in computing the state of the system neglecting the environment.
To do that, we introduce the density matrix  $\hat{\tilde{\rho}}_{SE}(t) = \ket{\tilde{\Psi}(t)}\bra{{\tilde{\Psi}(t)}}$.
Formally, we obtain its time evolution as
\begin{equation}
\hat{\tilde{\rho}}_{SE}(t) = \hat{\tilde{\rho}}_{SE}(0) - \ii \int_0 ^t \de t' \, \left[\hat{\tilde{H}}(t'),\hat{\tilde{\rho}}_{SE}(t') \right] 
\end{equation}
and therefore
\begin{equation}
\partial_t \hat{\tilde{\rho}}_{SE}(t) = - \ii \left[\hat{\tilde{\rho}}_{SE}(0),\hat{\tilde{H}}(t) \right] - \int_0 ^t \de t' \, \left[\hat{\tilde{H}}(t),\left[\hat{\tilde{H}}(t'),\hat{\tilde{\rho}}_{SE}(t') \right]\right] 
\end{equation}
We assume that the interaction term $\hat{\tilde{H}}(t)$ is too weak to create a significant correlation between the system and the bath. Furthermore, we also assume that any excitation of the environment due to its interaction with the system is dispersed on the infinitely many environment degrees of freedom. 
These approximations, collectively known as the \emph{Born approximation}, allow to consider that ${\hat{\tilde{\rho}}_{SE}(t)\simeq \hat{\tilde{\rho}}_{S}(t) \otimes \hat{\rho}_{E}}$, where $\hat{\rho}_{E}$
remain mostly unperturbed along the dynamics.
If $\omega_c \ll k_B T$, where $k_B$ is the Boltzmann constant and $T$ is the temperature of the bath, we can assume that the environment has always zero excitations in those degrees of freedom which can effectively couple with the system. Hence, $\hat{\rho}_{E}= \bigotimes_{\omega}\ket{0}\bra{0}$, where $\bigotimes_{\omega}$ indicates the tensor product over all the frequencies $\omega$.
We can now take the partial trace over the degrees of freedom of the environment, obtaining
\begin{equation}
\partial_t\hat{\tilde{\rho}}_S(t)=\partial_t \pTr{E}{\hat{\tilde{\rho}}_{SE}(t)}= \frac{\gamma}{2} \left(2\hat{a} \hat{\tilde{\rho}}_{S}(t) \hat{a}^\dagger -\hat{a}^\dagger \hat{a}  \hat{\tilde{\rho}}_{S}(t) - \hat{\tilde{\rho}}_{S}(t) \hat{a}^\dagger \hat{a}\right).
\end{equation}
Finally, by considering again the Schr\"odinger representation $\hat{\rho}_{S}= e^{-\ii \hat{H}_c t}\hat{\tilde{\rho}}_{S} e^{\ii \hat{H}_c t}$ we obtain the following master equation for a damped Harmonic oscillator:
\begin{equation}\label{Eq:Microscopic_derivation}
\partial_t\hat{\rho}_{S}(t)=-\ii \left[\hat{H}_c,\hat{\rho}_{S}(t) \right] +\frac{\gamma}{2} \left(2\hat{a} \hat{\rho}_{S}(t) \hat{a}^\dagger -\hat{a}^\dagger \hat{a}  \hat{\rho}_{S}(t) - \hat{\rho}_{S}(t) \hat{a}^\dagger \hat{a}\right).
\end{equation}
Indeed, \eqref{Eq:Microscopic_derivation} has the Lindblad form presented in \eqref{Eq:Lindblad_Master_Equation}.

\subsection{Measuring an optical cavity (or on infinity and the Zeno paradox)}
Let us consider two coupled systems, $\hat{a}$ representing an optical cavity, and $\hat{\sigma}_z$ representing a detector, whose Hamiltonian reads
\begin{equation}
\hat{H} = \omega \left(\hat{a}^\dagger\hat{a} + \frac{\hat{\sigma}_z}{2} \right)+ g \left( \hat{a}^\dagger \hat{\sigma}_- + \hat{a} \hat{\sigma}_+ \right).
\end{equation}
We periodically measure the second cavity with the following measurement instrument:
\begin{equation}
   \hat{M}_0 = \hat{\mathds{1}} \otimes \ket{0}\bra{0},  \hat{M}_1= \hat{\mathds{1}} \otimes \ket{0}\bra{1}, 
\end{equation}
That is, the measurement instrument $\mathcal{M}$ leaves unaffected the $\hat{a}$ cavity but if it detects one excitation in the detector, it destroys it.
The operators $\hat{M}_0$ and $\hat{M}_1$ are positive semi-definite (they have eigenvalues $1$ and $0$, or $0$, respectively), and
\begin{equation}
   \sum_{r} \hat{M}_r^\dagger \hat{M}_r = \hat{\mathds{1}} \otimes \left( \ket{0}\bra{0} +\ket{1}\braket{0|0}\bra{1} \right) =  \hat{\mathds{1}} \otimes \hat{\mathds{1}}.
\end{equation}
Thus, they define a POVM.
To each $\hat{M}_r$ we suppose that the measurement instrument will produce a different signal, say $0$ for $\hat{M}_0$ and $1$ for $\hat{M}_1$.

Let us initialize the system in $\ket{n,0}$ ($n$ photons in the cavity, no excitations in the detectors), and let us suppose that the system evolves for a time $\tau$ before it is measured.
The system wave function before the measurement is
\begin{equation}
\ket{\Psi(\tau)}= \cos(g \sqrt{n} \tau) \ket{n, 0} + \sin(g \sqrt{n} \tau) \ket{n-1, 1}.
\end{equation}
If at time $\tau$ we measure, we obtain, in average,
\begin{equation}\begin{split}
     \hat{\rho}'(\tau) &= \mathcal{M} \hat{\rho}(\tau) = \hat{M}_0  \ket{\Psi(\tau)} \bra{\Psi(\tau)}\hat{M}_0^\dagger + \hat{M}_1 \ket{\Psi(\tau)} \bra{\Psi(\tau)} \hat{M}_1^\dagger  \\ & =\cos^2(g \sqrt{n} \tau)\ket{n, 0}\bra{n,0} + \sin^2(g \sqrt{n} \tau)\ket{n-1, 0}\bra{n-1,0}.
     \end{split}
\end{equation}
We can now expand this expression, supposing that $\tau$ is very small, and obtain
\begin{equation}
    \hat{\rho}'(\tau) = \ket{n, 0}\bra{n,0} - g^2 \tau^2 n \left( \ket{n, 0}\bra{n,0} - \ket{n-1, 0}\bra{n-1,0} \right),
\end{equation}
and hence
\begin{equation}
\begin{split}
    \frac{\hat{\rho}'(\tau) -  \hat{\rho}(0)}{\tau} &=  g^2 \tau n \left(\ket{n-1, 0}\bra{n-1,0} - \ket{n, 0}\bra{n,0} \right) \\ 
    &= g^2 \tau \left( \hat{a} \ket{n, 0}\bra{n,0} \hat{a}^\dagger - \frac{\hat{a}^\dagger \hat{a} \ket{n, 0}\bra{n,0} + \ket{n, 0}\bra{n,0} \hat{a}^\dagger \hat{a}}{2} \right) \\ 
    & = g^2 \tau \left( \hat{a} \hat{\rho}(0) \hat{a}^\dagger - \frac{\hat{a}^\dagger \hat{a} \hat{\rho}(0) + \hat{\rho}(0) \hat{a}^\dagger \hat{a}}{2} \right) = g^2 \tau \DD[\hat{a}] \hat{\rho}(0)
    .
    \end{split}
\end{equation}
But since this is true for any $n$, we can conclude that any initial state $\hat{\rho}(0)$ would evolve with the same equation of motion, and similarly for any $\rhot$.

As one can now see, there is a manifest problem: although this equation looks like a Lindblad master equation for the optical damping, if we appropriately take the limit $\tau \to 0$ we do not observe any time evolution, and we conclude that $\partial_t \rhot =0$.
This is the famous Zeno paradox: the fact that the time evolution depends on $\tau^2$ upon continuous measurement makes it in impossible for a system to evolve.
So, how is it possible that the action of the environment, which is described as a similar measurement process, can actually produce dissipation?

The answer is simple: we did not take into account the fact that the environment has infinitely many degrees of freedom.
Indeed, a more accurate description needs to explicitly take into account that the environment is large, a fact that the previous model does not do.
There are two questions that one could ask: (i) How often does the environment ``measure'' (so how big is $\tau$)? (ii) how many modes are in the environment.
A very qualitative argument is the following.
The ``coherence time'' of a quantum system (i.e., for how long the environment will keep track of the fact that an excitation has been lost) is of the order of $\tau= A/L$, where $L$ is the size of the environment (very large); that is, the larger the environment, the shorter the correlation time.
The number of modes in the system is given (roughly) by the number of modes which have the same energy around the energy $\omega$ of the system within a volume $L$.
If we assume a particle-in-a-box type of system, and $E_n = \frac{n^2 \pi^2}{2m L^2}$, the number of eigenmodes $n(L)=B L$ is linearly dependent on $L$.
Therefore, we can re-write the previous equation as 
\begin{equation}
\begin{split}
    \frac{\hat{\rho}'(\tau) -  \hat{\rho}(0)}{\tau} = \sum_{j=1}^{n(L)} \frac{A g_j^2}{L} \tau \DD[\hat{a}] \hat{\rho}(0) \simeq  B L \frac{A  g^2}{L} \DD[\hat{a}] = AB g^2 \DD[\hat{a}]
    .
    \end{split}
\end{equation}
The constant $AB g^2\equiv \gamma$ is the dissipation rate of our system.

\subsection{Radiative damping via quantum jumps}

We saw previously that microscopic description of an optical resonator interacting with an environment at $T=0$ leads to $\jump=\hat{a}$ under the Born and Markov approximations.
Here, according to the previous discussion, we consider the same system and we make the hypothesis that the environment is made of nothing but a perfect photodetector.
First, the photodetector does not interact directly with the optical cavity, an therefore the Hamiltonian of the system is $\hat{H}_C$ defined in \eqref{Eq:Cavity_Alone}.
We consider that, if a photon is emitted into the environment, it is destroyed. Thus, the operator $\hat{M}_1$ must be of the form $\hat{M}_1 = \sqrt{\tau} \sum_n c_n \ket{n}\bra{n+1}$.
Let us define $c_0\equiv\sqrt{\gamma}$.
If we consider a cavity with one photon inside, the action of the operator $\hat{M}_1$, i.e., $\hat{M}_1 \ket{1}\bra{1}\hat{M}_1^\dagger = \tau \gamma \ket{0}\bra{0}$, describes the probability that in a time $\tau$ a photon is emitted.
We conclude that $\gamma$ is the decay rate of one photon inside the cavity (the lifetime is $1/\gamma$).
Now, if we have $n$ identical photons inside the cavity, we must require that this probability is $n$ times bigger.
We conclude that $c_n=\sqrt{\gamma} \sqrt{n}$.
But, by definition, $\hat{a}=\sqrt{n} \sum_n \sqrt{n}  \ket{n}\bra{n+1}$.
Hence, $\hat{M}_1= \sqrt{\tau} \hat{a}$, and therefore $\jump=\sqrt{\gamma}  \hat{a}$.
The time evolution of $\rhot $ is thus,
\begin{equation}\label{Eq:Measure_derivation}
\partial_t \rhot=-\ii \left[\hat{H}_C, \rhot\right] + \frac{\gamma}{2} (2\hat{a} \rhot \hat{a}^\dagger - \hat{a}^\dagger \hat{a} \rhot - \rhot \hat{a}^\dagger \hat{a}).
\end{equation}
Clearly, \eqref{Eq:Microscopic_derivation} and \eqref{Eq:Measure_derivation} coincide.

\section{Writing the Liouvillian as a matrix}
\label{sec:LiouvMatr}

As shown in Sec.~\ref{Sec:Form_of_Maps}, the Liouvillian  $\LL$ is a linear superoperator, since $\LL \left(\alpha \hat{\xi} + \beta \hat{\chi} \right) = \alpha \LL\hat{\xi} + \beta \LL \hat{\chi}$ for any  complex number $\alpha, \, \beta$ and any operator $\hat{\xi}, \, \hat{\chi}$.
Thus, it can be written as a matrix $\supmat{\LL}$.

The most naive way to write the Liouvillian matrix is to compute its matrix elements.
This procedure is straightforward, but it requires to explicitly write down an orthonormal basis for the density matrices and project the Liouvillian onto it.
For example, using the number basis $\ket{n}$, one constructs all the basis elements $H\otimes H$, which are of the form $\hat{\xi}_{(m,n)}=\ket{m}\bra{n}$.
We stress that $\hat{\xi}_{(m,n)}$ constitutes an orthonormal basis of $H\otimes H$, since any operator can be written as $\hat{\chi}=\sum_{n,m} c_{n,m} \hat{\xi}_{(m,n)}$ and $\Tr{\hat{\xi}_{(m,n)}^\dagger \hat{\xi}_{(m',n')}}=\braket{m|m'}\braket{n'|n}=\delta_{n,n'} \delta_{m,m'}$.
The matrix element of the Liouvillian are 
\begin{equation}\label{Eq:Matrix_Liouvillian_Slow}
\mathcal{L}_{(m,n),(p,q)} =\Tr{\hat{\xi}_{(m,n)}^\dagger \, \mathcal{L} \, \hat{\xi}_{(p,q)}} =\Tr{\hat{\xi}_{(m,n)}^\dagger \left(-\ii \left[\hat{H}, \hat{\xi}_{(p,q)}\right]+ \sum_\mu\mathcal{D}[\jump_\mu]\hat{\xi}_{(p,q)}\right)}.
\end{equation}

\subsection{Bosonic systems}

From a numerical point of view, the previously described procedure is very slow.
Indeed, for a cutoff $N$, it requires, in principle, to compute $N^4$ times \eqref{Eq:Matrix_Liouvillian_Slow}. 
A more efficient procedure passes through vectorisation of the operators, i.e. the linear transformation which converts a matrix into a column vector.

Let us consider a generic operator $\hat{\xi}$. 
The vectorisation of this matrix is:
\begin{equation}\label{Eq:vectorisation}
\hat{\xi}=\sum_{m,n} c_{m,n} \ket{m}\bra{n}\longrightarrow \vec{\xi}=\sum_{m,n} c_{m,n} \ket{m}\otimes \bra{n}^{\rm TR}=\sum_{m,n} c_{m,n} \ket{m}\otimes \ket{n^*},
\end{equation}
where the complex conjugate is a consequence of $\ket{n}=\bra{n}^\dagger=\left(\bra{n}^{\rm TR}\right)^*$.
In order to convert the Lindblad superoperator into its matrix form, we have to transform the right action superoperator $R[\hat{O}] \, \bigcdot =\bigcdot \, \hat{O}$ and the left one  $L[\hat{O}] \, \bigcdot=\hat{O} \, \bigcdot$ into their vectorised form $\supmat{R}[\hat{O}]$ and $	\supmat{L}[\hat{O}]$.
Let us start by the right action:
\begin{equation}\label{Eq:RightAction}
\begin{split}
	\supmat{R}[\hat{O}] \vec{\xi} & =\supmat{R}[\hat{O}] \sum_{m,n} c_{m,n} \ket{m}\otimes \ket{n^*}= 
	\overrightarrow{\xi \hat{O}}= \sum_{m,n} c_{m,n}  \, \ket{m}\otimes (\bra{n} \hat{O} )^{\rm TR}
	\\ & =\sum_{m,n} c_{m,n}  \, \ket{m}\otimes ( \hat{O}^{\rm TR}\ket{n ^*}) = (\mathds{1}\otimes \hat{O}^{\rm TR}) \vec{\xi}. 
\end{split}
\end{equation}
In the same way, we have:
\begin{equation}\label{Eq:LeftAction}
\supmat{L}[\hat{O}] \vec{\xi}= (\hat{O}\otimes \mathds{1}) \vec{\xi}. 
\end{equation}

From the result of Eqs.~\eqref{Eq:RightAction}~and~\eqref{Eq:LeftAction}, we can eventually write any Liouvillian $\LL= - \ii \left[\hat{H}, \bigcdot\right] + \gamma/2 \; \mathcal{D}[\jump]$  (for simplicity with only one jump operator $\jump$) under the form
\begin{equation}
	\begin{split}
	\supmat{\mathcal{L}}&= -\ii \left(\supmat{L}(\hat{H}) - \supmat{R}(\hat{H}) \right)+\left(2 \supmat{L}(\jump) \supmat{R}(\jump^\dagger) - \supmat{L}(\jump^\dagger \jump) -\supmat{R}(\jump^\dagger \jump) \right) \\ &
	= -\ii \left((\hat{H}\otimes \mathds{1})  - (\mathds{1}\otimes \hat{H}^{\rm TR})\right) + \frac{1}{2} \left(2 \jump\otimes \hat{\Gamma^*}-\jump^\dagger \jump\otimes \mathds{1} -  \mathds{1} \otimes  \jump^{\rm TR} \jump^* \right).
	\end{split}
\end{equation}



\end{document}